%% file: main.tex
\documentclass[sigconf]{acmart}

\AtBeginDocument{%
  }

\setcopyright{acmlicensed}
\copyrightyear{2018}
\acmYear{2018}
\acmDOI{XXXXXXX.XXXXXXX}
\acmConference[Conference acronym 'XX]{Make sure to enter the correct
  conference title from your rights confirmation email}{June 03--05,
  2018}{Woodstock, NY}
\acmISBN{978-1-4503-XXXX-X/2018/06}

\input{math_commands}

\usepackage{graphicx}
\usepackage{hyperref}
\usepackage{url}
\usepackage{graphicx}
\usepackage{verbatim} 
\usepackage{subcaption} 
\usepackage{amsthm}
\usepackage{svg}
\usepackage{makecell}

\newtheorem{definition}{Definition}

\usepackage{enumitem}  
\usepackage{comment}
\usepackage{algpseudocode}
\usepackage{algorithm}

\usepackage{graphicx}
\usepackage{adjustbox}
\usepackage{multicol}

\usepackage{tcolorbox}
\tcbuselibrary{breakable}
\tcbset{
  width=0.9\textwidth,
  halign=justify,
  center,
  breakable,
  colback=white    
}
\usepackage{mdframed}

\usepackage{xspace}
\newcommand{\company}{Microsoft\xspace}
\usepackage{verbatimbox}

\usepackage[dvipsnames]{xcolor}
\usepackage{pgfkeys}
\usepackage{tcolorbox}
\colorlet{LightLavender}{Lavender!40!}
\usepackage{fontawesome}

\begin{document}

\title{A Holistic Framework for Automated Configuration Recommendation for Cloud Service Monitoring}


\author{Anson Bastos}
\affiliation{%
  \institution{Microsoft}
  \country{India}
  }
\email{ansonbastos@microsoft.com}
\author{Shreeya Venneti}
\affiliation{%
  \institution{Microsoft}
  \country{India}
  }
\author{Anjaly Parayil}
\affiliation{%
  \institution{Microsoft}
  \country{India}
  }
\email{aparayil@microsoft.com}
\author{Ayush Choure}
\affiliation{%
  \institution{Microsoft}
  \country{India}
  }
\author{Chetan Bansal}
\affiliation{%
  \institution{Microsoft}
  \country{United States}
  }
\email{chetanb@microsoft.com}
\author{Rujia Wang}
\affiliation{%
  \institution{Microsoft}
  \country{United States}
  }
\email{rujiawang@microsoft.com}



\begin{abstract}
    Reliability of large-scale cloud services is critical for user satisfaction and business continuity. Despite significant investments in reliability engineering, production incidents remain inevitable, often leading to customer impact and operational overhead. In large cloud companies, multiple services are deployed across regions necessitating robust health monitoring systems.
    However, the current monitor configuration process is manual, largely reactive and ad hoc, resulting in gaps in coverage and redundant alerts. In this paper, we present a comprehensive study of monitor creation in \company, identifying key components in the existing process. We further design a modular recommendation framework that processes the graph structured service entities to suggest optimal monitor configurations. 
    Through extensive experimentation on historical data and user study of recommendations for production services
    at \company, we demonstrate the efficacy of our approach in providing relevant recommendations for monitor configurations. 
\end{abstract}




\maketitle

\input{Introduction}
\input{ProblemFormulation}

\input{Analysis}

\input{UnifiedFramework}

\input{Results}
\input{Conclusion}

\bibliographystyle{ACM-Reference-Format}
\bibliography{sample-base}


\end{document}

%% file: math_commands.tex
\usepackage{amsmath,amsfonts,bm}

\def\eqref#1{equation~\ref{#1}}

\def\1{\bm{1}}

\def\ve{{\bm{e}}}

\def\vx{{\bm{x}}}

\DeclareMathAlphabet{\mathsfit}{\encodingdefault}{\sfdefault}{m}{sl}
\SetMathAlphabet{\mathsfit}{bold}{\encodingdefault}{\sfdefault}{bx}{n}

\def\gE{{\mathcal{E}}}

\def\gG{{\mathcal{G}}}

\def\gV{{\mathcal{V}}}



%% file: Introduction.tex
\section{Introduction}

Large scale cloud service providers such as Amazon, Microsoft, Google etc. run thousands of services over complex environments. At \company, the hyper-scale cloud infrastructure supports over 5,000 services deployed across more than 60 regions, serving hundreds of millions of users globally. Ensuring the continuous availability of these services is critical to sustaining customer satisfaction and preserving business revenue \cite{surianarayanan2019}. Despite extensive investments in reliability engineering, production incidents and failures remain unavoidable, often resulting in customer impact, financial losses and requiring substantial engineering effort for detection, diagnosis and mitigation. Consequently, early detection and resolution of such incidents are vital to minimizing user disruption and reducing operational costs. To this end, service providers engage in continuous health monitoring of services, aiming to proactively identify and address issues before they affect end users.

The lifecycle of an incident generally begins with detection, followed by triaging to the concerned stakeholders for resolution who then diagnose the incident and apply some mitigation so that the service becomes functional. This is generally followed by a detailed root causing of the incident and working towards a longer term solution for example by making necessary code changes. Incidents could either be detected by automated watchdogs called \emph{monitors} or reported by customers. The latter is undesirable as it results in loss of revenue and affects the reputation of the company. Further manual incident reporting delays detection which increases service downtime. In this paper we focus on the detection phase of an incident's lifecycle, specifically by monitors. 

The current methodology for configuring monitors primarily follows a trial-and-error paradigm. Service owners define monitors based on their understanding of the service architecture and service-level objectives but do not consider the intricate relation between existing monitor configurations and the past incident detection performance. Additionally, monitors are often revised or newly introduced in response to production incidents. However, this approach has notable limitations. First, it is inherently reactive, which means critical monitors may be absent until an incident reveals the gap. Second, it frequently leads to the creation of redundant monitors, resulting in excessive alert noise and unnecessary operational overhead.

\textbf{Related Works:} In recent years, numerous empirical studies have explored the challenges associated with monitoring and incident resolution in cloud services \cite{10.1145/2670979.2670986, 10.1145/2987550.2987583, 10.1145/3317550.3321438, 10.5555/3691825.3691834, 9276598, 10.1145/3542929.3563482, 10.1145/3236024.3236030, 10.1145/3368089.3417055, 9505089}, as well as the practical difficulties in defining and adhering to Service Level Objectives (SLOs) \cite{10.1145/3317550.3321432, 8831196_1, 8831196_2}. These prior works generally assume that the appropriate monitors are already identified. \cite{metric_paper_icse} studied what aspects of a service should be monitored and which metrics are most relevant for assessing the performance, efficiency, and reliability of cloud-based systems. In similar vein, \cite{ibm_metric_selection} studies the problem of selecting the optimal set of metrics for monitoring. However, these works only focus on a component of the monitor creation process and to the best of our knowledge there is no prior work that studies and implements an end to end configuration recommendation pipeline for cloud service monitors.

This work addresses a critical and underexplored problem in software engineering: the \emph{automated configuration of service monitors in large-scale cloud systems}. In modern software engineering practice, especially within cloud-native environments, monitoring and incident detection are foundational to ensuring system reliability, availability, and maintainability. Despite the ubiquity of monitoring tools, the configuration of monitors remains largely manual, reactive, and error-prone.
We identify the following key software engineering challenges addressed by our work:
1) \textbf{Monitor Configuration Complexity}: Service owners must manually define monitors without systematic guidance, often leading to gaps in coverage, redundant alerts, and delayed incident detection. This problem is exacerbated by the scale and heterogeneity of cloud services.
2) \textbf{Lack of Intelligent Support for Monitor Design}: Existing tools offer limited support for recommending configurations based on historical data, service context, or domain-specific patterns. There is a need for an \emph{intelligent, data-driven framework} that can assist engineers in \emph{configuring effective monitors}.

\textbf{Contributions:} In this paper, we systematically study the monitor creation process for 3600 accounts/services across \company. We begin by the analysis of the individual components that are used in the creation of a monitor. Guided by the empirical study, we build modules that recommend the best configuration for each component. We further present a framework that unifies the results from the individual modules and provides a concise configuration recommendation for the monitors of the respective services
in a low code setup.
In summary, we make the following contributions:
\begin{itemize}[leftmargin=*]
    \item We present an empirical study and analysis of the individual components of the monitor creation pipeline.
    \item We build models to recommend component-wise configurations.
    \item We unify the results of individual modules into a single framework for service owners to create monitors.
    \item We conduct extensive evaluation and production user study of our method to understand the performance efficacy.
\end{itemize}

%% file: ProblemFormulation.tex
\section{Problem Setting}

In this section, we establish the formal context for monitoring in a cloud service setting. We begin by introducing a set of definitions that outline the foundational concepts within the service monitoring domain. Next, we present the specific operational context at \company in which this study is situated. Finally, we articulate the monitor configuration recommendation problem and detail our proposed formulation.

\subsection{Definitions}
We define the below terms in the context of micro-services:

A \underline{\emph{metric}} is defined as a time-series entity that is generated through the utilization of a resource and is therefore inherently linked to that resource. Implicit in this definition is a mathematical expression that characterizes the elements of the time series. A typical example is RAM utilization observed on a machine hosting a service. We denote individual metrics using the lowercase letter $m$, and for a given metric $m$, the associated resource is denoted as $r_m$. These metrics could be emitted across various attributes called \underline{\emph{dimensions}}, which can be thought of as the granularity along which the data has to be aggregated. In the running example, the RAM utilization could be monitored across a single machine or a group of machines in the cluster or region and so on. The level and granularity across which a metric is aggregated and emitted is decided by the service owner depending on the monitoring needs and is referred to as the \emph{dimension selection} problem.

\underline{\emph{Expressions}} are mathematical forms that are tied to certain metrics as defined by the user. For example in the case of $\text{{\tt RAM utilization}} = \frac{\text{{\tt Raw RAM utilization in MBs}}}{\text{Total RAM capacity in MBs}}$, we could define the corresponding expression (for {\tt RAM utilization}) using the metric of {\tt Raw RAM utilization in MBs} and the maximum available capacity.

\underline{\emph{Alerting logic}} refers to a set of anomaly detection rules that operate over metrics and serve as triggers for alert creation. For instance, a rule such as {\tt RAM utilization > 85\% for the past 20 time steps} would raise an alert when the memory is likely to become a performance bottleneck. Each rule is associated with a severity level and is tied to a specific \emph{expression}. We denote individual alerting logic statements using the symbol $A$, with the corresponding expression represented as $E_a$.

A \underline{\emph{monitor}} is formally defined as a collection of tuples, each comprising a metric, the dimensions to aggregate across for each metric and an associated alerting logic with corresponding expressions. A monitor triggers an alert of a specific severity level when any of its constituent alerting logic rules of the same severity are activated. We denote an individual monitor by the capital letter $M$. The monitor $M$ consists of $k$ such tuples, represented as $\{m_i,d_i,E_i, A_i\}_{1 \leq i \leq k}$, where each tuple corresponds to a distinct alerting logic statement.

\begin{figure}[t]
    \centering   \begin{subfigure}[b]{0.5\textwidth}
     \centering
\includegraphics[width=\linewidth]{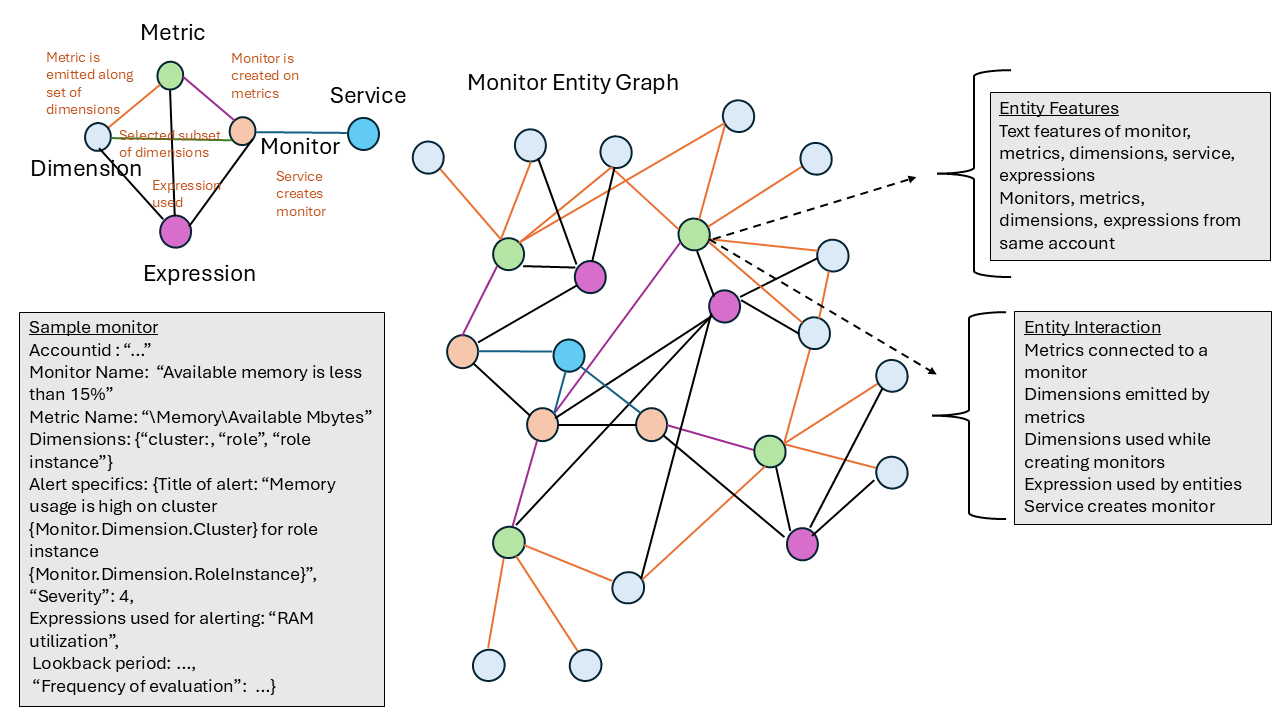}
        \caption{}
           \label{subfig:formulationa}
    \end{subfigure}%
    
    \begin{subfigure}[b]{0.5\textwidth}
         \centering
\includegraphics[width=\linewidth]{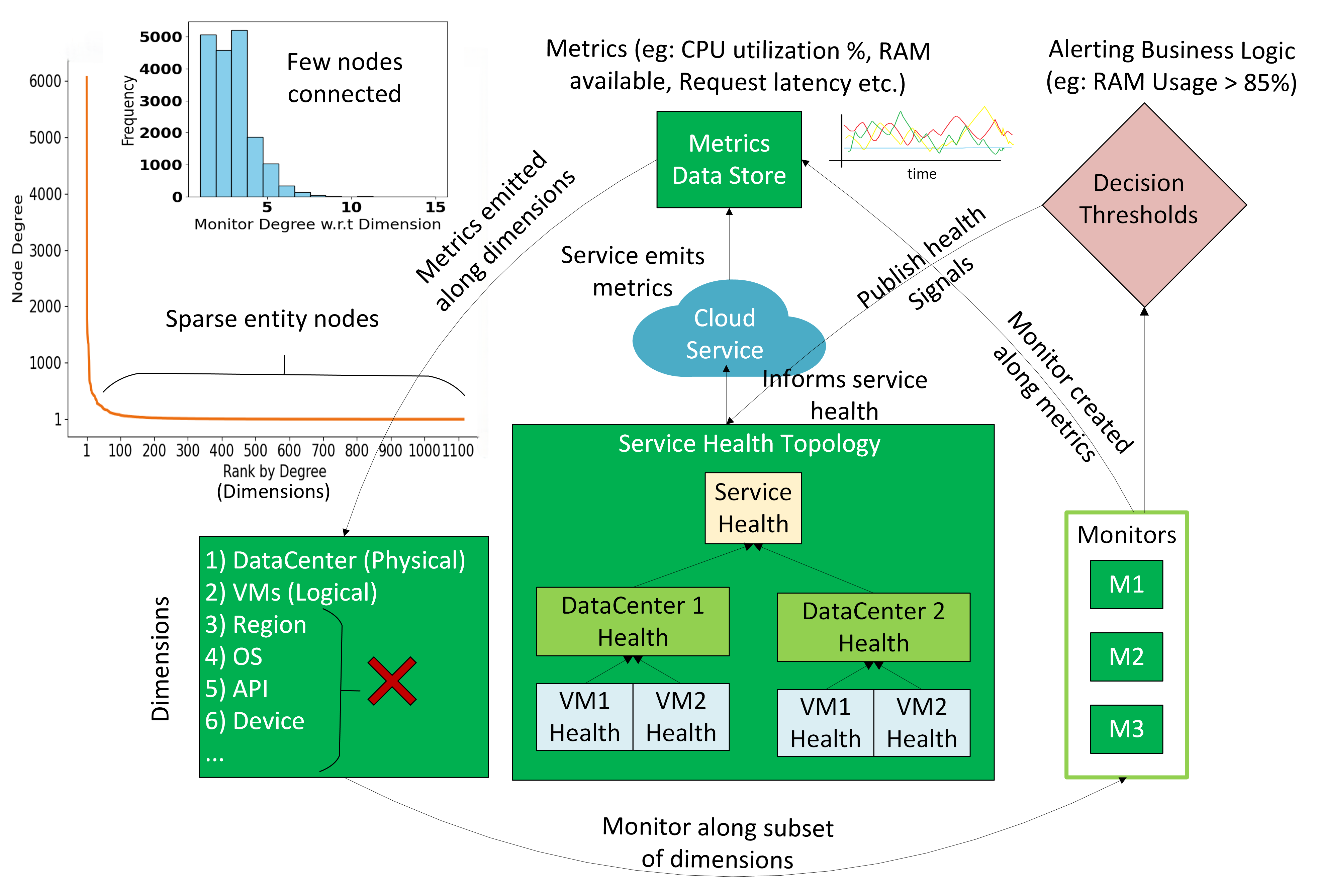}
    \caption{}
           \label{subfig:formulationb}
    \end{subfigure}
\caption{ System Overview: a) Monitor entity graph: Nodes represent the monitors, metrics, and dimensions in a cloud setting. Each node contains text features and interacts with their neighboring nodes using link communication and b) The overall service health where the cloud service emits metrics along many dimensions and only few are being used for monitoring health. For eg. here, the service health depends on health of each datacenter using the service which in turn depends on the health of the individual VMs in the datacenter. The alerting conditions are applied over the defined expressions on metrics and are decided by the business need.
}
\label{fig:formulation}
\end{figure}

\subsection{Monitor Entity Data Structure} \label{monitor_entity_graph}

Having looked at the individual entities of a monitoring system, we now define the graph network formed by the interconnection of these components.
\begin{definition}{(Monitor Entity Graph):}
We represent the data as a heterogeneous graph \(\displaystyle \gG =(\displaystyle \gV, \displaystyle \gE) \) where \(\displaystyle \gV=\{\displaystyle \gV_s \cup \displaystyle \gV_m \cup \displaystyle \gV_d \cup \displaystyle \gV_k \cup \displaystyle \gV_{exp}\}\) represents the set of nodes with \(\displaystyle \gV_s, \displaystyle \gV_m, \displaystyle \gV_d, \displaystyle \gV_k, \displaystyle \gV_{exp}\) denoting services, monitors, dimensions, metrics and expressions respectively and \(\displaystyle \gE=\{\displaystyle \gE_{sm} \cup \displaystyle \gE_{md} \cup \displaystyle \gE_{kd} \cup \displaystyle \gE_{mk} \cup \displaystyle \gE_{mexp} \cup \displaystyle \gE_{kexp} \cup \displaystyle \gE_{dexp}\}\) represents the set of edges, capturing following types of relationships:
1) \(\displaystyle \gE_{sm}\): ``service creates monitor'', 2) \(\displaystyle \gE_{md}\): ``monitor associated with dimension'', 3) \(\displaystyle \gE_{kd}\): ``metric has dimension'', 4) \(\displaystyle \gE_{mk}\): ``monitor emits metric'' 5) \(\displaystyle \gE_{mexp}\): ``monitor evaluates expression''. 6) \(\displaystyle \gE_{kexp}\): ``metric used in expression'' 7) \(\displaystyle \gE_{dexp}\): ``dimension used in expression''
\end{definition}

The graph structure \(\gG\) encapsulates the intricate relationships in the data that are necessary for recommending the configurations. By explicitly modeling different entity types and their interconnections, we enable the model to learn domain-specific patterns. Each node in the graph, denoted as \(v \in \gV\), has a unique initial representation given by \(\vx_v \in \mathbb{R}^d\), where \(d\) is the dimension of the embedding space. The vector \(\vx_v\) is the concatenation of two types of features: \emph{ontology} features, which are domain-specific textual attributes of the entity (e.g., metric name, dimension name, monitor name, expressions name, related service), and learned embeddings, which are trainable embeddings that capture the entity's role in the graph structure (e.g., co-occurrence with another node). Figure \ref{fig:formulation} gives a broad overview of the monitor entity graph along with the overall service health monitoring system, for a better understanding of the application context. In the figure, we see the service health depends on health of each datacenter (dimension 1) using the service which in turn depends on the health of the individual VMs (dimension 2) in the datacenter and the monitor is configured accordingly with the alert threshold set for the "$\text{{\tt RAM utilization (in \%)}}$" expression (defined over "$\text{{\tt Raw RAM utilization in MBs}}$" metric) being ">85\%".

\noindent \textbf{Notations:} 
We briefly state the notations used for clarity. $\gG = (\gV, \gE)$  is the heterogeneous monitor entity graph with nodes $\gV$ and edges $\gE$. The nodes consist of services ($v_s \in \gV_S$), monitors ($v_m \in \gV_M$), metrics ($v_k \in \gV_K$), dimensions ($v_d \in V_D$), expressions ($v_e \in \gV_E$). The alerting logic, which is the combination of thresholding rules over the expressions are denoted as $a$. $\gE_{ij}$ denotes the edge type from node $i$ to $j$. $\mathbf{h}_v$ is the feature vector (embedding) of node $v$ obtained from the \emph{ontology} (textual attributes) of the node and $\mathcal{N}(v)$ is its neighborhood in the graph.

\subsection{Monitor Configuration Recommendation}
In this section we describe the problem of the configuration recommendation for monitors. 
We are interested in recommending the following entities: 1) Metrics, 2) Dimensions, 3) Expressions, 4) Alert conditions. As we have seen in the previous sections, the metrics are the time series data that are being monitored over specific resources. These are emitted along certain dimensions that the service owner is interested. Expressions then use these metrics in a mathematical form that evaluates to the desired measure. The alert condition then sets the criteria of business rules for raising an alert in case the thresholds are violated in the defined manner. Thus the framework is divided into these main modules, each for recommending the above entities in a sequential manner as we shall see below and in the further sections. The recommendations are then surfaced to the user in a natural language interface to provide coherent reasoning and recommendations.

Since we deal with heterogeneous graph structured data, in order to recommend the entities we use a graph based framework. We pose the problem of recommendation as an entity ranking problem. 
Thus, we formulate the problem of selecting the optimal monitor configuration as a \emph{recommendation problem over the monitor entity graph}, which we formally state as follows:
\begin{definition}[Problem formulation]
Leverage  entity features \{\(\displaystyle \vx_v: v \in    \displaystyle \gV \) \}, link features $\{ \displaystyle \ve_{v_1v_2} \in  \displaystyle \gE\}$ and monitor entity graph, $\displaystyle \gG$ to learn entity representations that facilitate relevant sequential recommendations of metric, dimension and expression nodes ($v_k, v_ d, v_e \in \gV$) for a given tuple of ($v_s, v_m \in \gV$), ($v_s, v_m, v_k \in \gV$), ($v_s, v_m, v_k, v_d \in \gV$) respectively and to generate alerting conditions over the expressions for the selected ($v_s, v_m, v_k, v_d, v_e \in \gV$).
\end{definition}

%% file: Analysis.tex
\section{Empirical Study}\label{empirical_study}

In this section, we study the different aspects of the monitor entity graph (cf. \S \ref{monitor_entity_graph}) and discuss the findings. 
The experimental setup is over the graph constructed using the service data from $\sim$ 3600 accounts of \company consisting of $\sim$ 60k historical monitors, $\sim$ 14K metrics, $\sim$ 7K dimensions, $\sim$ 27K expressions and $\sim$ 2M alerting conditions. 
We study the following: 1) Sparsity in the nodes / entities in the monitor entity graph. This will help to understand the interaction characteristics of the network and help in designing the recommendation model. 2) Relation between ontology (textual) features of the nodes. Analyzing this aspect will help to understand if the textual features can be used to deduce similar entities 3) What are the mathematical forms used in the expressions? This will help to decide on the method to select the appropriate expression for a monitor. 4) Whether the monitoring status (i.e. to monitor a metric or not) and alert conditions and thresholds can be effectively deduced from the metric timeseries data. This is an interesting aspect as, if the timeseries data could be a good predictor of the conditions we could bypass the graph data and simply use the metric's timeseries. If not then we would need to explore the induction of the alert conditions from semantically similar monitors from the historical network data. This brings us to the final aspect to study namely 5) Whether there is any correlation between the alert conditions and the features of semantically similar entities?
Based on the analysis and derived observations in this section, we formulate the monitor configuration recommendation framework in the subsequent sections.

\begin{figure}[h!]  
\centering
\begin{subfigure}[b]{0.22\textwidth}
       \centering
\includegraphics[width=1.0\linewidth]{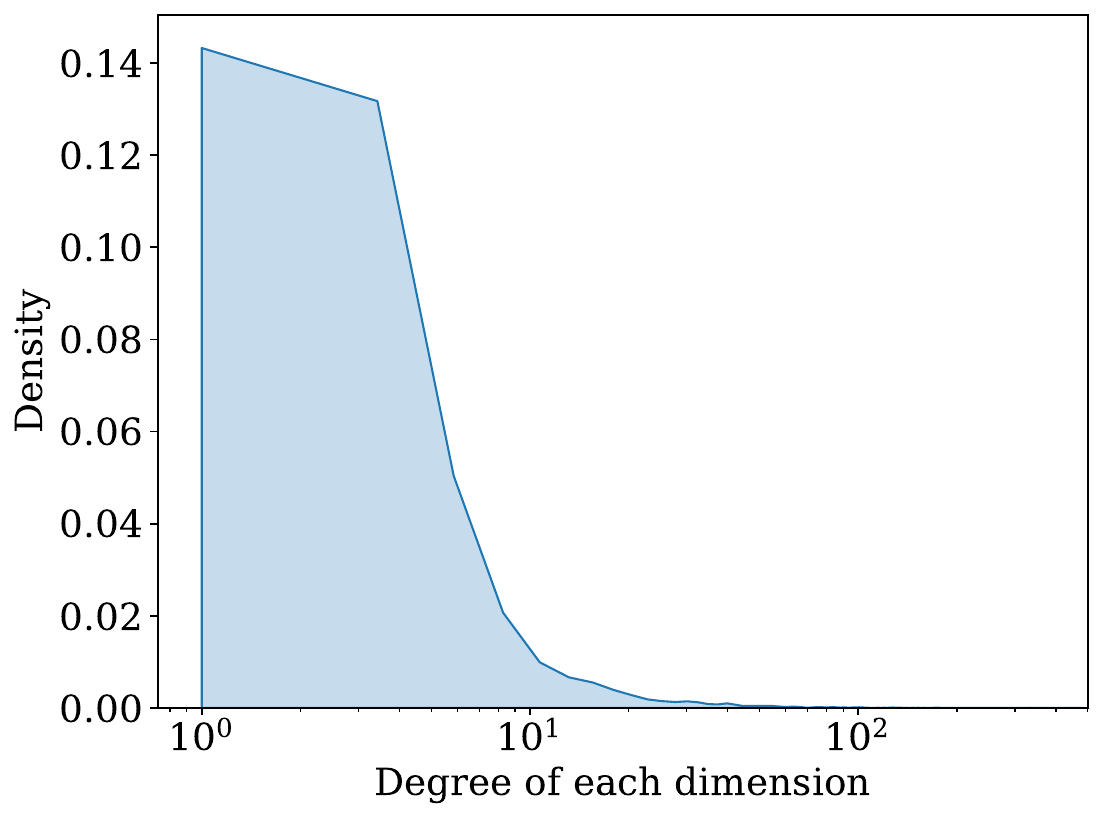}
\caption{}
\label{subfig: dimpopularity2}
\end{subfigure}%
\begin{subfigure}[b]{0.22\textwidth}
       \centering
\includegraphics[width=1.0\linewidth]{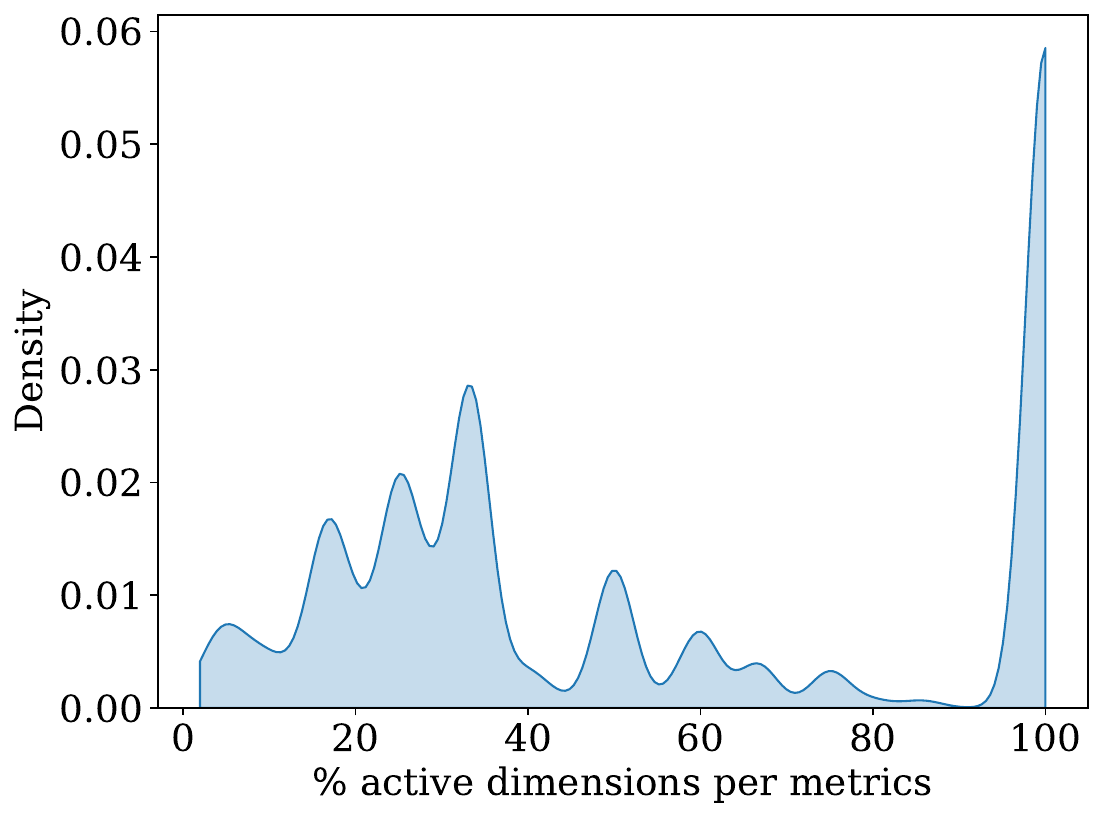}
\caption{}
\label{subfig: dimpopularity3}
\end{subfigure}
    \caption{
    Characteristics of the Monitor Entity Graph: a) Distribution of degree associated with dimensions based on the metric-to-dimension links, b) Distribution of the percentage of dimensions selected from the set of all dimensions along which the metric is emitted. 
    }
    \label{fig: dimpopularity}
\end{figure}

\noindent \textbf{Sparsity in the nodes / entities in the monitor entity graph:} \autoref{fig: dimpopularity} illustrates the sparsity characteristics of the monitor entity graph. 
\autoref{subfig: dimpopularity2} shows the distribution of degree associated with dimensions based on the metric-to-dimension links (i.e., the number of metrics to which each dimension is connected). 
Similarly, \autoref{subfig: dimpopularity3} shows the distribution of dimension degree (i.e. dimensions actually selected) based on the monitor-to-dimension interactions from the monitor entity graph. The likelihood test ratio indicates a resemblance to long-tailed distributions, indicating that the majority of monitors (94\%) do not need to aggregate the metrics along all dimensions along which they are emitted. 
Thus post selection of metrics we need to select the optimal subset of dimensions along which to emit the timeseries data for monitoring.
Thus the problem of selecting an optimal subset of dimensions on which to monitor is relevant.

\begin{tcolorbox}[width=0.5\textwidth,
                  boxsep=0pt,
                  left=2pt,
                  right=2pt,
                  top=2pt,
                  arc=0pt,
                  boxrule=0pt,leftrule=1pt,
                  colback=LightLavender
                  ]
 \faLightbulbO
\emph{
Observation 1: 
The ``monitor entity'' graph exhibits 
activity sparsity. Although many dimensions are associated with metrics, only a subset of them is used to aggregate the metric while raising an alert.}
\end{tcolorbox}

\begin{figure}
    \centering
    \includegraphics[width=0.7\linewidth]{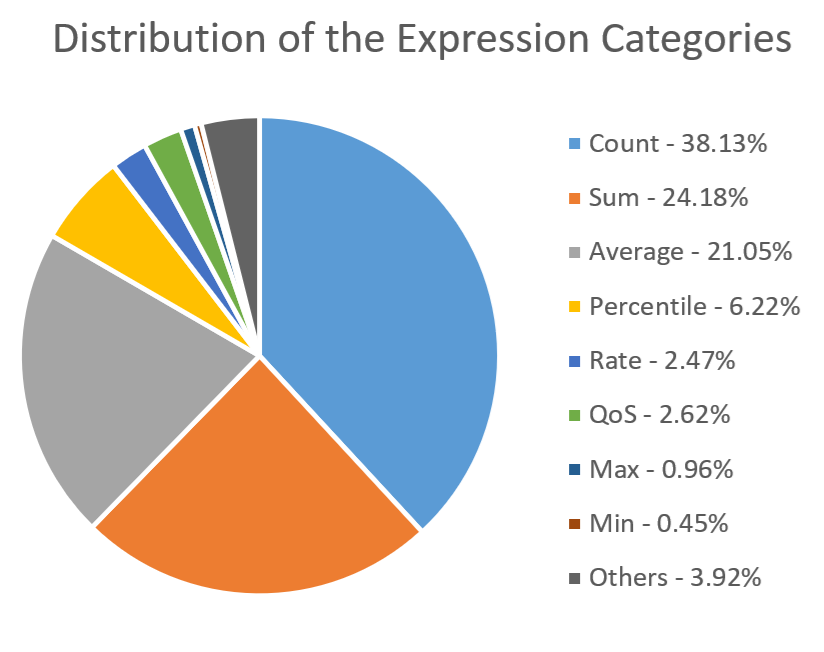}
    \caption{Study of the categorization of the Expressions used by the monitors. We can observe that most expressions (~83\%) use either of count, sum or average as the mathematical form to aggregate the metrics data.}
    \label{fig:expression_types}
\end{figure}

\noindent \textbf{Relation between ontology (textual) features of the nodes:} Next, we study the features associated with the entities to gauge the effect on recommendations. We analyze the text features associated with the monitor entity graph (c.f. \autoref{subfig:formulationa}) and its effect on dimensions associated with the monitors.
\begin{figure}[h!]  
\centering
\begin{subfigure}[b]{0.22\textwidth}
       \centering
\includegraphics[width=\linewidth]{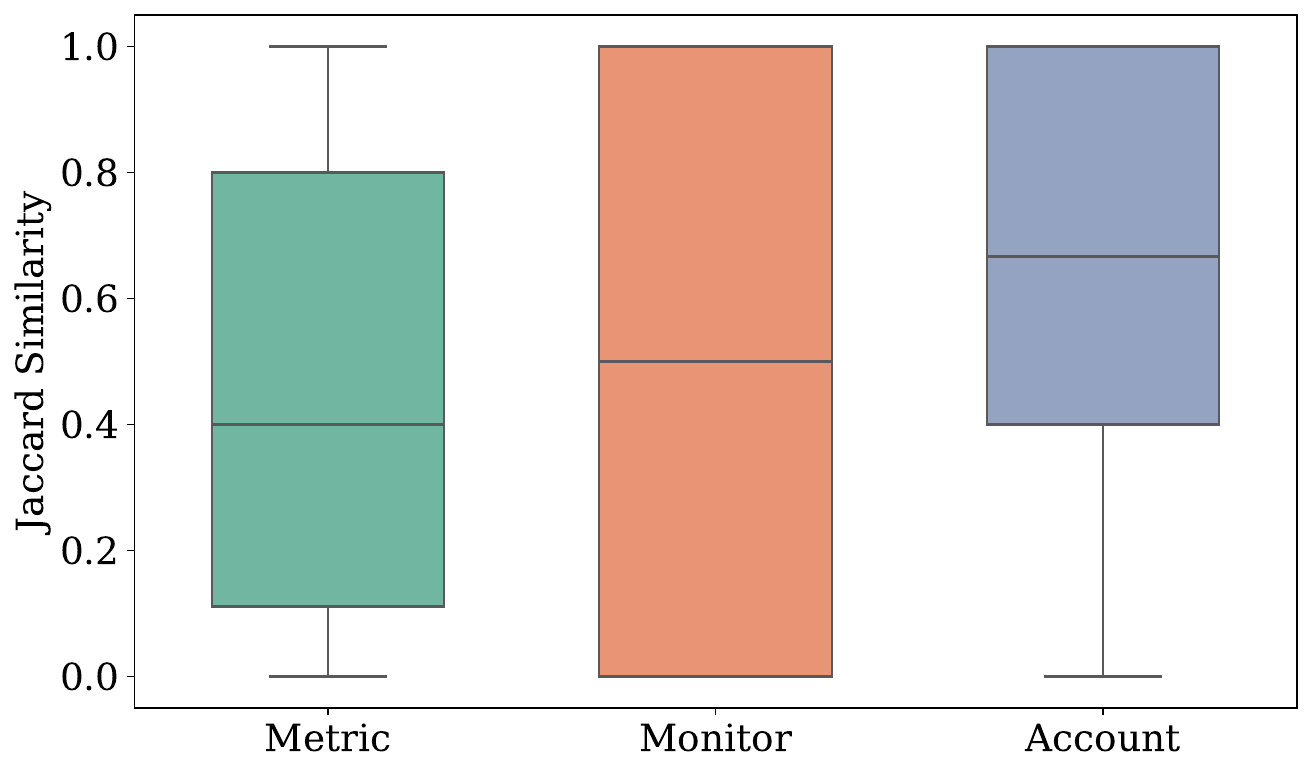}%
\caption{}
\label{subfig: graphanalysis1}
\end{subfigure}
\begin{subfigure}[b]{0.22\textwidth}
       \centering
\includegraphics[width=\linewidth]{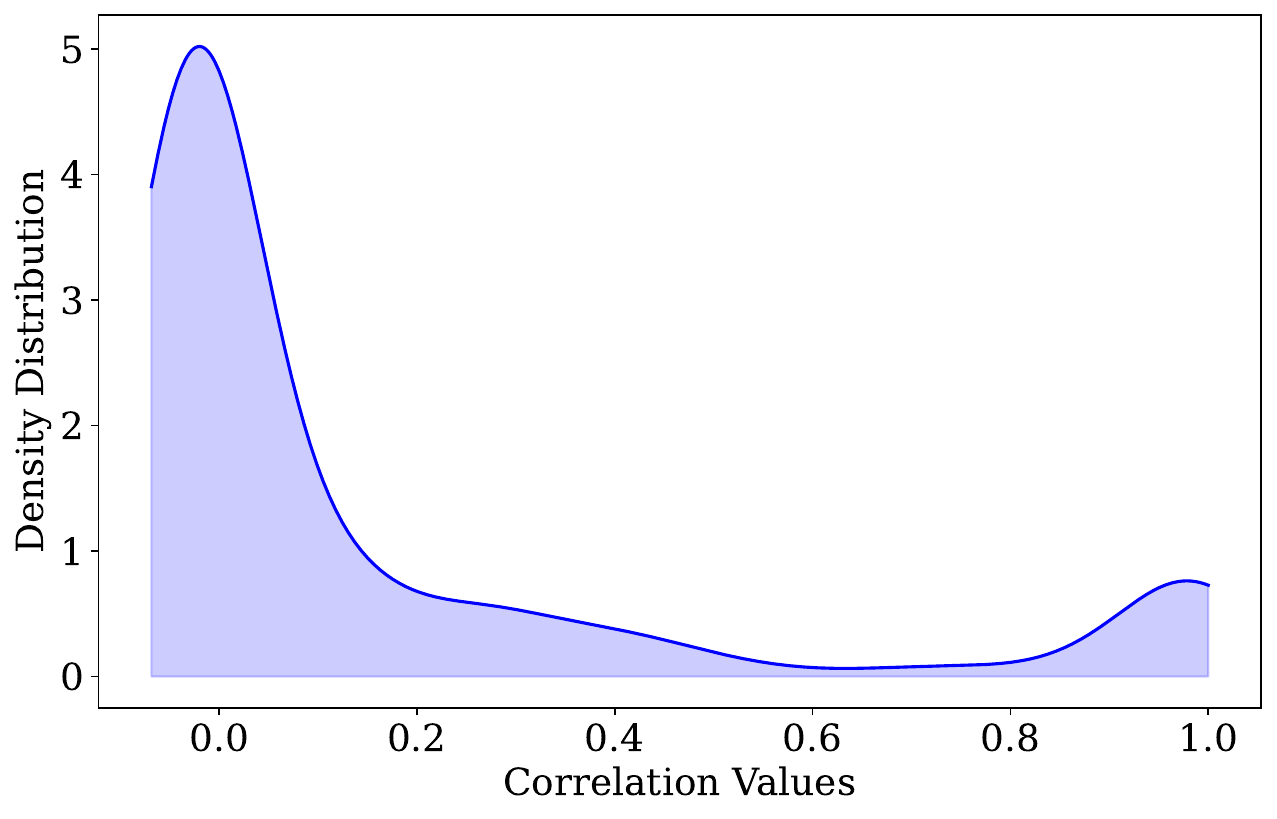}
\caption{}
\label{subfig: graphanalysis2}
\end{subfigure}
    \caption{a) Variation in jaccard similarity of set of dimensions associated with monitors with similar metric, monitor names, and same service account, and b) Distribution of pairwise correlation between dimensions.
    }
    \label{fig: graphanalysis}
\end{figure}
\autoref{subfig: graphanalysis1} illustrates the distribution of Jaccard similarity across sets of dimensions linked to monitors that demonstrate high cosine similarity (greater than 0.8) based on various textual attributes. These attributes include monitor names, metric names, and the corresponding service account. Textual embeddings are derived using the "E5" model, a general-purpose embedding technique trained via contrastive learning \citep{wang2022text}. The observed Jaccard similarity patterns differ depending on whether the similarity is based on metric names, monitor names, or shared service accounts. Notably, dimension similarity associated with similar monitor names exhibits greater variance. Additionally, \autoref{subfig: graphanalysis2} presents the density distribution of pairwise correlations among dimensions connected to a monitor. This plot reveals a bimodal distribution, indicating the existence of two distinct groups with differing correlation behaviors. The second peak reflects the frequent co-occurrence of a specific subset of dimensions.
\begin{tcolorbox}[width=0.5\textwidth,
                  boxsep=0pt,
                  left=2pt,
                  right=2pt,
                  top=2pt,
                  arc=0pt,
                  boxrule=0pt,leftrule=1pt,
                  colback=LightLavender
                  ]
 \faLightbulbO
\emph{
Observation 2: 
Monitors associated with the same service account, similar metrics, or naming conventions often share common dimensions, though the extent of overlap varies. Additionally, dimension correlations form two distinct clusters—one positively correlated and one largely uncorrelated—highlighting the need for node representations that reflect both feature similarity and diverse correlation patterns.
}
\end{tcolorbox}
\noindent \textbf{Mathematical forms used in the expressions:} In this study we shall explore the categories of expressions used in the alerting logic. These expressions assume some mathematical forms such as summing, averaging and so on over the metrics' timeseries data. This study informs us what are the major types of expressions that are being evaluated by existing monitors and how to select the expression for a new monitor. From \autoref{fig:expression_types}, we observe the following major types of expressions in current use by the monitors: Count, Sum, Average, Percentile, Rate, QoS, Max, Min. These cover more than 95\% of the expressions. 
\begin{tcolorbox}[width=0.5\textwidth,
                  boxsep=0pt,
                  left=2pt,
                  right=2pt,
                  top=2pt,
                  arc=0pt,
                  boxrule=0pt,leftrule=1pt,
                  colback=LightLavender
                  ]
 \faLightbulbO
\emph{
Observation 3: 
As most of the expressions use similar mathematical forms as their counterpart monitors, we treat the problem of recommending expressions as finding similar expressions from existing monitors after learning over the monitor entity graph.}
\end{tcolorbox}
\noindent After deducing the expression form, the actual evaluation would be performed over the metrics of the concerned monitor.

\begin{figure}[h!]  
\centering
\begin{subfigure}[b]{0.24\textwidth}
       \centering
\includegraphics[width=1.0\linewidth]{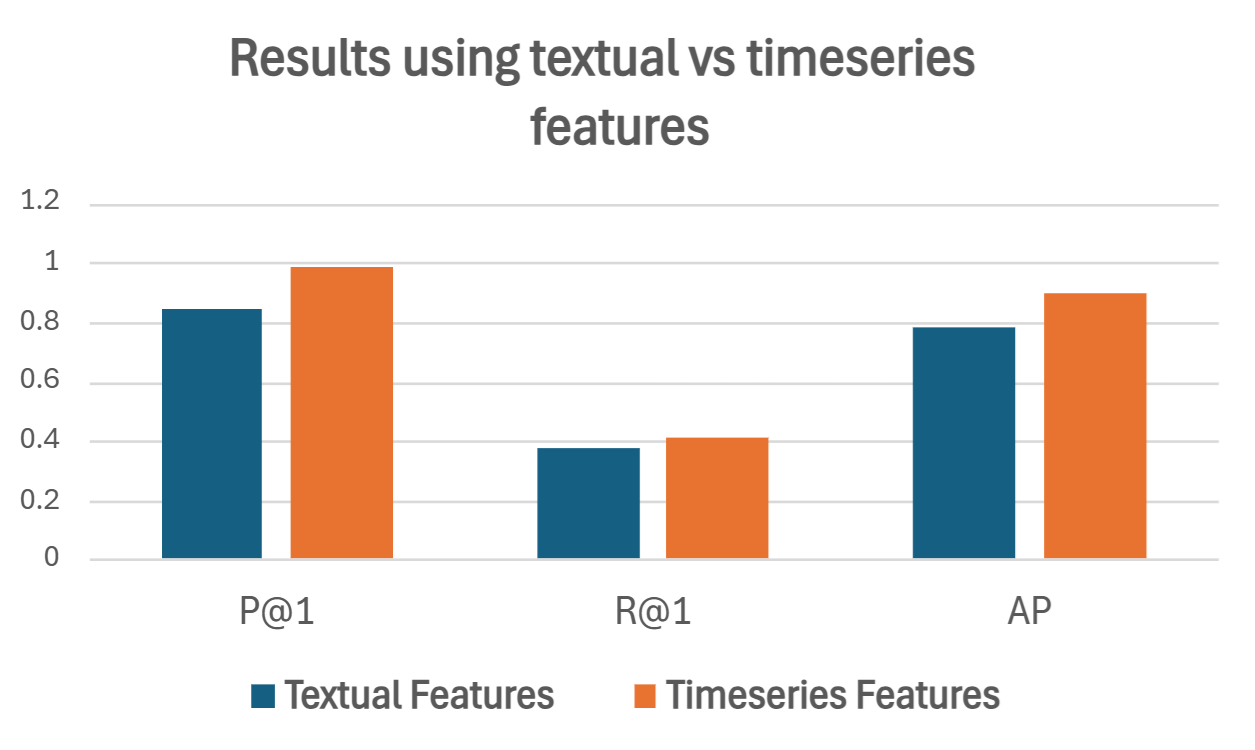}
\caption{}
\label{subfig: metric_ts_analysis1}
\end{subfigure}%
\begin{subfigure}[b]{0.24\textwidth}
       \centering
\includegraphics[width=1.0\linewidth]{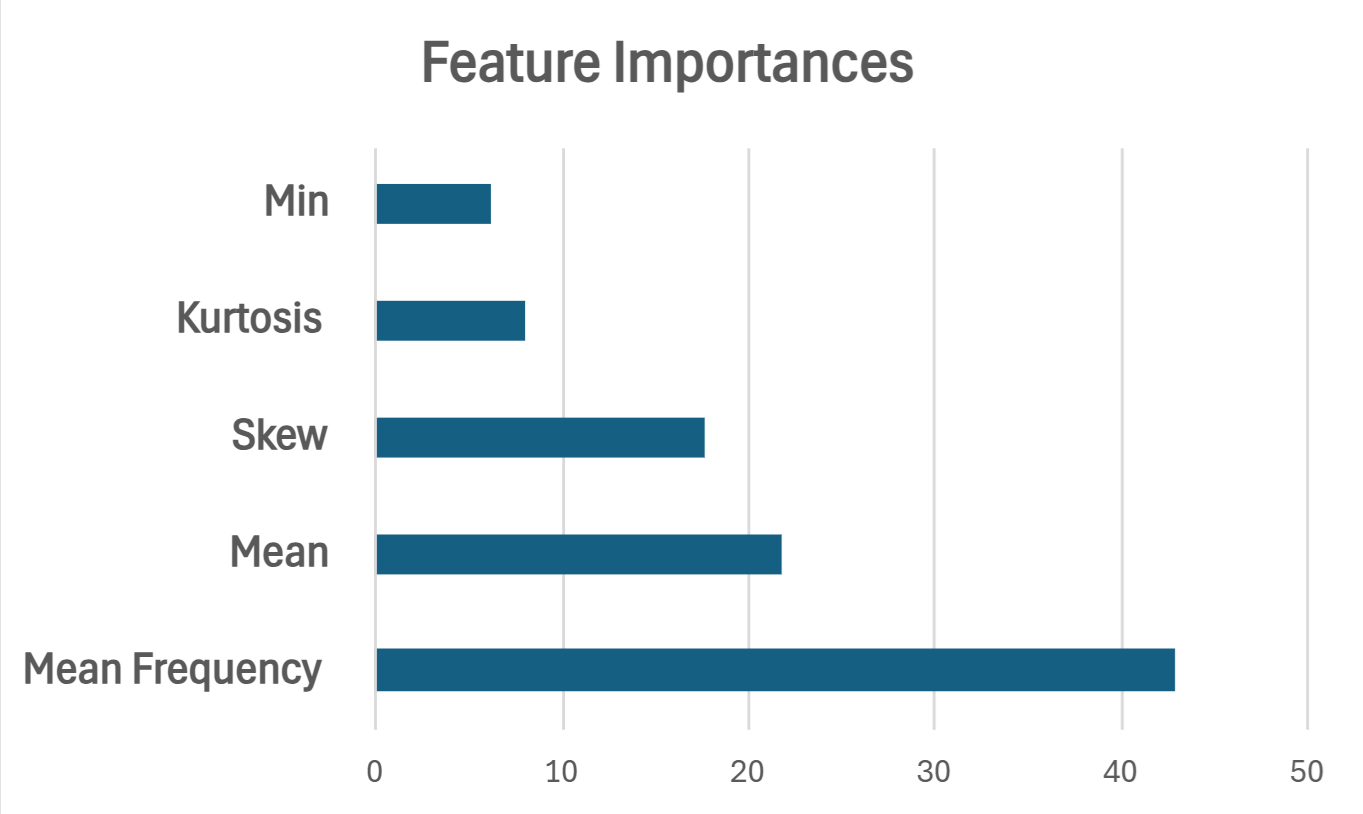}
\caption{}
\label{subfig: metric_ts_analysis2}
\end{subfigure}
    \caption{
    Analysis of the importance of time series features on predicting the monitoring status of the metric.
    }
    \label{fig: metric_ts_analysis}
\end{figure}

\noindent \textbf{Monitoring status and alerts from timeseries:} In the next study we explore if the timeseries data of metrics can be used to predict i) the monitoring status (whether the metric should be monitored or not) along with ii) the alert conditions. Note one may naively decide to monitor all the available metrics; however there is a cost associated with monitoring each metric and hence it becomes important to select which metric should be monitored. The hypothesis for i) is that most metrics are static or the signals do not vary and only those metrics with some activity (varying frequency information) are important for monitoring. For this study we extract some statistical features from the timeseries data and perform a random forest \cite{random_forest_Breiman2001} based classification to answer the RQ. The features used are as follows: minimum, maximum, mean, median, mode, skew, kurtosis, mean frequency, maximum frequency. We compare the timeseries feature with the textual feature. Figure \ref{fig: metric_ts_analysis} presents the results from which we can see that we can decide with high precision whether or not a metric has to be monitored using timeseries features. Moreover, the performance due to timeseries features are statistically significant (p<0.05 using two-sided Student's test) than using textual features. The feature importance shows the frequency feature is of highest importance validating the hypothesis that frequency information or variation in signal of metrics is a good predictor of whether the metric should be monitored. For ii) we build a simple linear regression model with the above timeseries features to predict the threshold. 
Unfortunately we find the $R^2$ score to be $< 0.1$. 
This is expected since the timeseries data may either not contain anomalous values in the analyzed range or the raw values may not be indicative of the business rules. 
\begin{tcolorbox}[width=0.5\textwidth,
                  boxsep=0pt,
                  left=2pt,
                  right=2pt,
                  top=2pt,
                  arc=0pt,
                  boxrule=0pt,leftrule=1pt,
                  colback=LightLavender
                  ]
 \faLightbulbO
\emph{
Observation 4: 
The timeseries data can help to select the metrics for monitoring with high precision but is not a good indicator of the alert thresholds.
The study informs us that the timeseries data can be used to select the metrics but not to decide the alert thresholds.}
\end{tcolorbox}

\begin{figure}[h!]  
\centering
\begin{subfigure}[b]{0.22\textwidth}
       \centering
\includegraphics[width=1.0\linewidth]{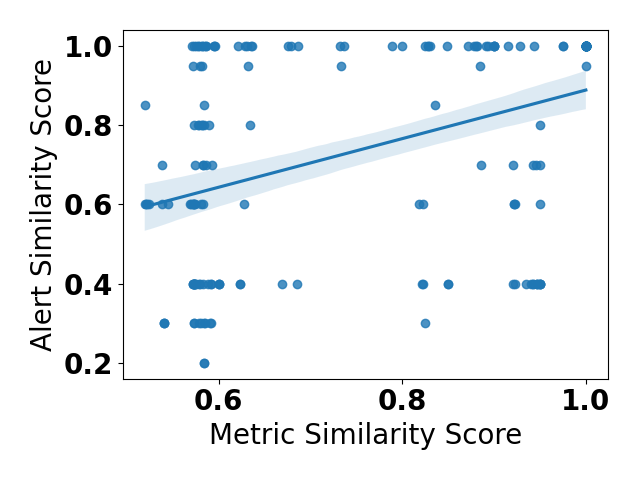}
\caption{}
\label{subfig: metric_similarity_analysis1}
\end{subfigure}%
\begin{subfigure}[b]{0.22\textwidth}
       \centering
\includegraphics[width=1.0\linewidth]{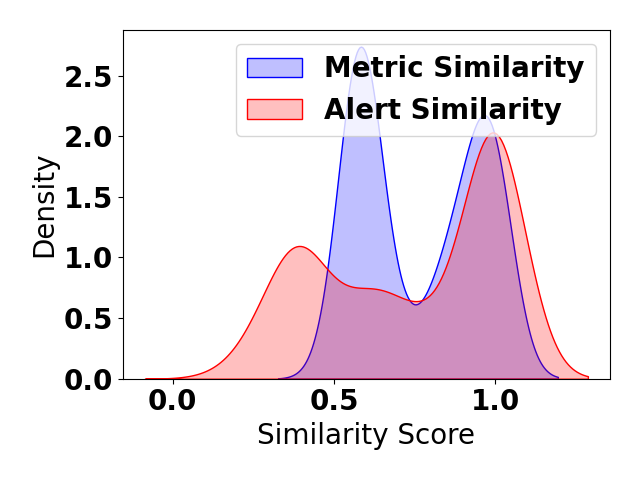}
\caption{}
\label{subfig: metric_sim_analysis2}
\end{subfigure}
    \caption{
    Study showing the relation between alert conditions and the feature based similarity. Figure (a) shows the correlation between the similarity score of the alert conditions (predicted vs actual) and the similarity score of the features of the entities (metrics). The correlation (0.41, p<0.05) indicates that the alert condition of similar metrics also show high similarity. Figure (b) shows the distribution of the similarity scores displaying a similar multimodal distribution.
    }
    \label{fig: metric_sim_analysis}
\end{figure}

\noindent \textbf{Correlation between the alert conditions and features of semantically similar entities:} In \autoref{fig: metric_sim_analysis} we study the relation between alert conditions and features of entities. Specifically, we computed the similarity scores between the metrics and the alert conditions of the similar metrics and compute the correlation between them. The similarity for the metric features is obtained using textual similarity (in this case, Jaccard similarity) and time series similarity (shapelet comparisons \cite{monitor_assistant}) and the two scores averaged out. For alert condition similarity, we use an LLM to evaluate the closeness based on defined criteria
(see \S \ref{unified_framework_section} for details).
\autoref{fig: metric_sim_analysis} shows good correlation (0.41) between the similarity scores of metric features and alert conditions. 
\begin{tcolorbox}[width=0.5\textwidth,
                  boxsep=0pt,
                  left=2pt,
                  right=2pt,
                  top=2pt,
                  arc=0pt,
                  boxrule=0pt,leftrule=1pt,
                  colback=LightLavender
                  ]
 \faLightbulbO
\emph{
Observation 5: 
We find a high correlation between the similarities of the metrics and alerts conditions. This indicates that finding similar metrics can help to deduce a similar set of alerting conditions.}
\end{tcolorbox}
\noindent These findings motivate the approach for alert generation in \S \ref{unified_framework_section}.

%% file: UnifiedFramework.tex
\section{A Unified Framework for Monitor Configuration Recommendation} \label{unified_framework_section}

\begin{figure*}
    \centering
    \includegraphics[width=0.9\linewidth]{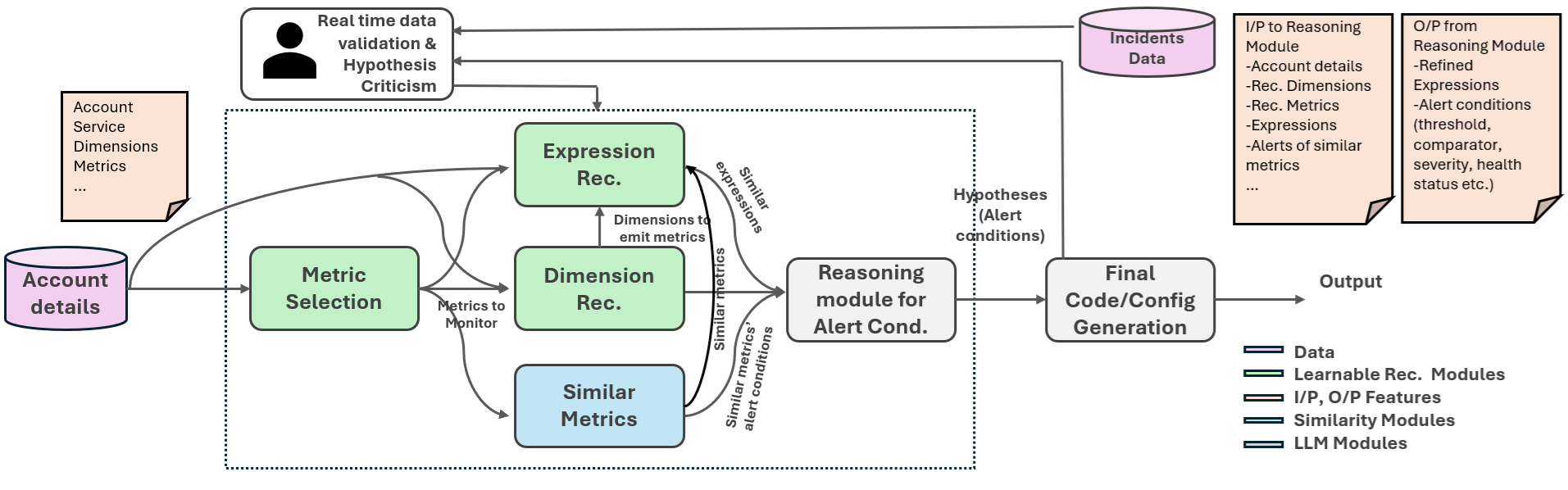}
    \caption{Overview of the monitor configuration recommendation framework. The input to the system is the account/service details, metrics created by the services and the dimensions associated with those metrics. The metric selection module first selects the subset of metrics to monitor followed by the dimension and expression recommendation modules recommending the dimensions to emit along and the expressions to evaluate respectively. The metric similarity module gathers similar metrics from historical data. The reasoning module (LLM) finally combines all the recommendations and data from the previous modules to generate the final alert conditions. The output module generates the final code needed to create the monitor in an automated manner. Once the monitor is created using the recommended configuration, the system gathers feedback from incidents and human subject matter experts (SMEs) to refine the recommendations in a continuous loop.}
    \label{fig:unified_fw_overview}
\end{figure*}

In this section, we describe in detail the individual modules composing the recommendation system and see how we unify the individual recommendations into a single coherent interface providing feedback to the system.
As seen in the previous sections we build modules for recommending metrics, dimensions, expressions and then use these to generate the alert conditions. 
The entire framework for the automated monitor configuration recommendation system is provided in figure \ref{fig:unified_fw_overview}. 
The system takes as input the details of the account or service, the metrics generated by the services, and the dimensions linked to those metrics. Inspired by \emph{observation 1, 2} in $\S$ \ref{empirical_study}, we select a subset of the metric and dimensions that are relevant for the monitor. Initially, the metric selection module identifies a relevant subset of metrics to monitor. This is followed by the dimension and expression recommendation modules, which suggest appropriate dimensions to emit and expressions to evaluate. Note while (mathematical) expressions could also be generated, however from \emph{observation 3} we find that the majority of expressions are reused between monitors. Thus we reason that ranking between metrics should work better and confirm the hypothesis in $\S$ \ref{results_sec} (cf. fig. \ref{fig: main_results}c). Further, A metric similarity module retrieves historically similar metrics to inform the configuration of alerting conditions of similar metrics/monitors. The reasoning module, powered by 
an LLM, 
synthesizes the outputs from preceding modules to formulate the final alerts (inspired by \emph{observations 4,5}). 
The output module then generates the code required to instantiate the monitor. The LLM is used to perform semantic similarity, format the alerts and provide reasoning for the service owner.
\autoref{fig:rec_framework} depicts an overview of the recommendation modules. 
We look at each module in detail below.
\subsection{Selecting subset of Metrics for monitoring}\label{metric_rec}
For the metric selection module we begin with the following features: service name and description, dependent service details, metrics available for the service and the dimensions associated for the metrics. We compute the sentence embeddings for the text features of each of the component using a sentence embedding model such as E5 \cite{wang2022text}. For the set of metrics, dimensions, dependent services etc., once the text features are obtained, we compute the set embeddings using a method similar to DeepSets \cite{deepsets_NIPS2017_f22e4747}. Specifically we use the mean pooling strategy as the aggregation method. We also experimented with attention mechanism but in our experiments we did not find much benefit. The final set embeddings are concatenated and passed to a 2 layer MLP \cite{MLP_Rumelhart1986} with ReLU activation followed by a classification head. The objective is to classify whether the metric should be monitored or not. Thus we pose it as a 2 class classification problem and use the binary cross entropy loss for optimization as follows
\begin{equation}\label{l_bce}
    \mathcal{L}_{bce} = \sum_{i} y_i \log(\hat{y_i}) + (1-y_i) \log(1-\hat{y_i})
\end{equation}
where $y_i = \{0,1\}$ is the indicator of whether the metric is to be monitored and $\hat{y_i}$ is the probability of monitoring the metric as predicted by the module. The data is split in a time based manner for training and evaluation. One caveat with this method is it assumes a global separation exists for the metrics to be monitored. However this may not always be true. For example, there may be a metric which is important to monitor for one service but not for another. Thus we also need to consider the local dynamics of the metrics given a service/account. In order to do this we consider the decision boundary in the vicinity of the metric given the account information. Specifically, we consider the metrics in the latent space (i.e. the network output) and perform a k nearest neighbors search and take the majority voting of the monitoring status of the metrics in the vicinity. To define the optimization objective we sample positive and negative metrics examples from historical data for the given account and compute a contrastive loss as below
\begin{equation}\label{l_contrastive}
    \mathcal{L}_{contrast}(e_k) = \sum_{k^+,k^-} \left( \lVert e_k^+ -e_k \rVert - \lVert e_k^- -e_k \rVert \right) + \left[ \gamma + \lVert e_{cent}^+ - e_{cent}^- \rVert \right]_+
\end{equation}
In the above equation, the loss is computed for the given metric embedding ($e_k$), $k^+$ denotes the positive metrics (i.e. of similar monitoring status) and $k^-$ are the negative (dissimilar monitoring status) pairs with $e_k$, $e_{cent}$ represents the euclidean centroid of a set of embeddings, $\gamma$ is a positive margin (set to 1 empirically) and $[.]$ denotes the positive value of the term (else 0). To minimize the loss the first term would try to have the positive metrics closer and negative ones farther away in embedding space whereas the second term would keep the positive and negative embeddings distinct within a margin. Since this loss works at the local decision boundary surrounding a metric for a given account, it captures the local dynamics. In order to jointly optimize the global and local dynamics we combine equations \ref{l_bce} and \ref{l_contrastive} as below 
\begin{equation}\label{l_joint}
    \mathcal{L}_{joint} = \alpha \mathcal{L}_{bce} + (1-\alpha) \mathcal{L}_{contrast}
\end{equation}
where $\alpha$ is a hyperparameter, set to 0.5.

\textbf{Selecting similar metrics}: As seen in section \ref{empirical_study}, the alert conditions of similar metrics are also similar and provide valuable insights into the alert conditions needed for the concerned metric. Thus we extract metrics similar to the recommended metric. In order to do this we adopt a two fold approach. Specifically as studied in section \ref{empirical_study}, we make use of both the ontology as well as the metric timeseries features in order to obtain the similar metrics. 
For the textual features, we select top metrics based on closeness in the embedding space. These are then input to an LLM to provide a score from 0 to 1 (no match to exact match) for the similarity of the metric descriptions. 
For the timeseries features we extract shapelets from the timeseries data similar to \cite{monitor_assistant}. Shapelets help in computationally efficient comparison of the timeseries data by extracting relevant shapes of interest in the data such as spikes, ramps etc. At the implementation level, we use \cite{fast_shapelets} to extract the shapelet and then compare shapelets of two metrics' timeseries using \cite{shape_based_distance}. Finally we combine the ontology and timeseries similarity by taking a weighted average of the two to get a similarity score between metrics. 
The alert conditions (along with dimensions, expressions etc.) of the similar metrics are used to inform the final alert conditions as we shall see in section \ref{alert_cond_gpt}.

\subsection{Graph based Recommendation}
Once we decide the metrics to monitor as seen in the previous section we now have to decide the dimensions along which the metrics would emit and the expressions to evaluate. We will compute these by learning over the monitor entity graph \(\displaystyle \gG \) defined in section \ref{monitor_entity_graph}.
The graph structure, \(\displaystyle \gG \), encapsulates both the intrinsic attributes of each entity and its relational context within the graph. 
We detail the message passing mechanism and optimization objectives followed by a context of the downstream recommendation tasks.

\begin{figure}[h!]
    \centering

\begin{subfigure}[b]{0.47\textwidth}
       \centering
\includegraphics[width=\linewidth]{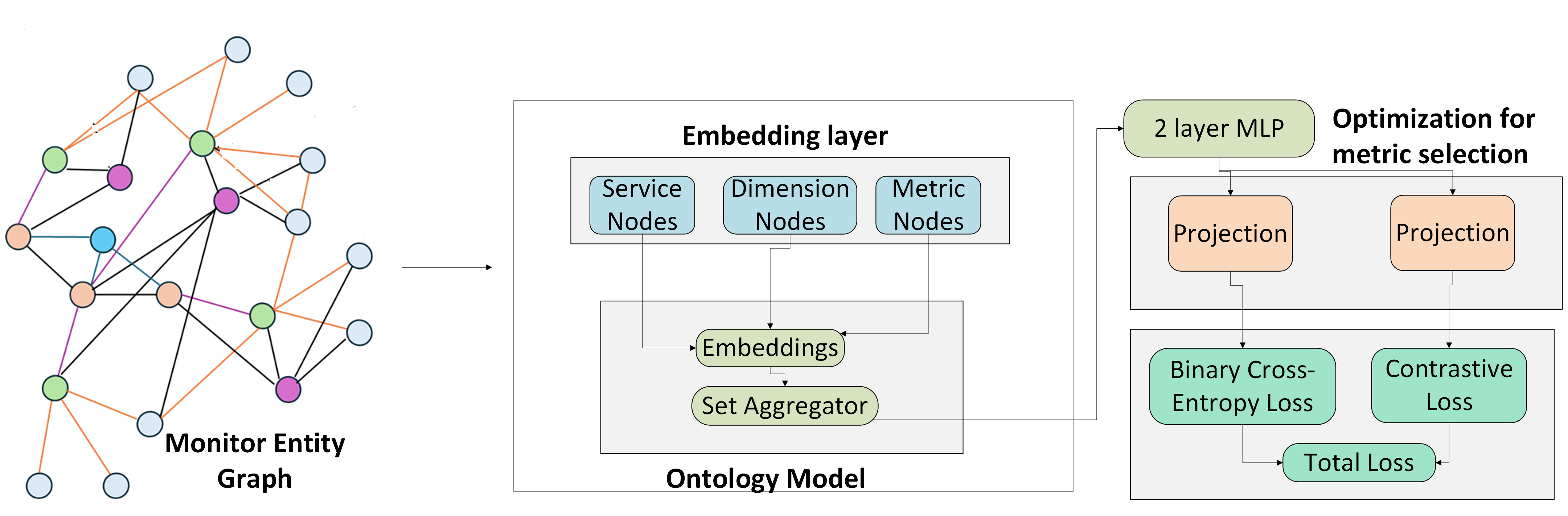}%
\caption{}
\label{subfig: rec_framework1}
\end{subfigure}
\begin{subfigure}[b]{0.47\textwidth}
       \centering
\includegraphics[width=\linewidth]{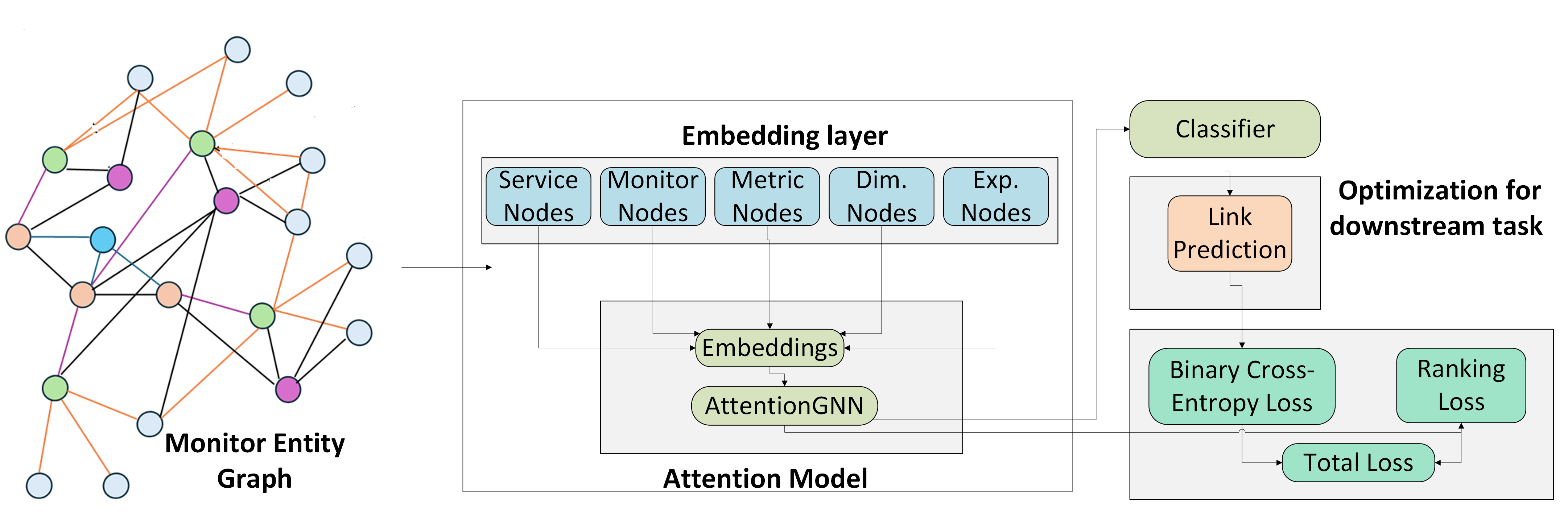}%
\caption{}
\label{subfig: rec_framework2}
\end{subfigure}

    \caption{ The overall architecture of the recommendation modules. Figure a shows the metric recommendation module which uses a 2 layer MLP over the set aggregated service features. Figure b shows the architecture used for the dimension/expression recommendation modules.  The framework applies an enhanced transformer-based graph convolution. }
    \label{fig:rec_framework}
\end{figure}

\subsubsection{Message Passing (MP) Mechanism}
In the context of heterogenous graphs for cloud monitoring and recommendation, our 
framework employs an effective message passing mechanism designed to capture complex relationships and contextual information (fig. \ref{fig:rec_framework}).
The multi-head attention enhanced message passing approach leverages edge-specific transformations and heterogeneous neighborhood aggregation. 
The key components of the mechanism are:

\textbf{Multi-Head Attention}:
A multi-head attention mechanism is employed to enable the model to simultaneously capture diverse aspects of node relationships. 
The attention weights \(\alpha\) are computed as follows:
$\alpha_{ij}=\frac{(q_i \cdot k_j)}{\sqrt{d_h d_o}}$,
    where \(q_i\) is the query vector for node \(i\), \(k_j\) is the key vector for node \(j\), \(d_h\) is the number of attention heads, and \(d_o\) is the output dimension per head.
    The attention weights are then normalized using softmax:
    $\alpha_{ij} = \frac{\exp(\alpha_{ij})}{\sum_{k \in \mathcal{N}(i)} \exp(\alpha_{ik})}$ 
    where \(\mathcal{N}(i)\) is the set of neighbors of node \(i\).
      
 \textbf{Edge-Aware Message Transformation}: 
 Each transformed message integrates the node’s feature representation with its corresponding attention weight, ensuring sensitivity to the specific type of edge. The transformed message \(m_{ij}\) from node \(j\) to node \(i\) is computed as $m_{ij} = \alpha_{ij} v_j $, where \(v_j\) is the value vector of node \(j\). The final aggregated message for node \(i\) is $m_i = \sum_{j \in \mathcal{N}(i)} m_{ij}$. The node features are updated as: \newline 
 \begin{equation}\label{eq_mha}
      x_i^{(l+1)} = \sigma\left(W^{(l)} \cdot \text{CONCAT}\left(x_i^{(l)}, m_i^{(l)}\right)\right)
 \end{equation}
 where \(x_i^{(l)}\) is the feature vector of node \(i\) at layer \(l\), \(W^{(l)}\) is the learnable weight matrix for layer \(l\), \(\sigma\) is the ReLU activation function and CONCAT is the concatenation operation.

\textbf{Incorporating Meta-Paths to Capture Global Properties}:
Although the attention mechanism described in the previous section effectively captures local graph structure, prior research has shown that it primarily excels at learning homophilic patterns—where neighbouring nodes share similar attributes—while struggling to model long-range dependencies \cite{balcilar2021analyzing, bastos2022how}. 
To address this limitation, we explicitly incorporate global information by introducing random walk paths originating from the target nodes. We adapt the meta path based method similar to \cite{sehgnn,HetGNN}.

\textbf{Training Objective}:
To address the multifaceted nature of our recommendation setting, we define a composite loss function that integrates multiple optimization goals (\autoref{fig:rec_framework}). This loss function comprises two essential components: (1) Binary Cross-Entropy, (2) TOP1-max Ranking Loss.

\textbf{Binary Cross-Entropy Loss}: The binary cross-entropy (BCE) loss is a fundamental loss  which is defined as:
\begin{equation}
    \mathcal{L}_{\text{BCE}} = -\frac{1}{N} \sum_{i=1}^N \left[ y_i \log(\hat{y}_i) + (1 - y_i) \log(1 - \hat{y}_i) \right]
\end{equation}
where \(y_i\) is the true label and \(\hat{y}_i\) is the predicted probability. 
The BCE loss ensures basic prediction accuracy.

\textbf{TOP1-max Ranking Loss}: To enhance the ranking quality of recommendations, we adopt the TOP1-max ranking loss introduced in \cite{hidasi2018recurrent}. 
It is formally expressed as:
\begin{equation}
    \mathcal{L}_{\text{TOP1-max}} = \sum_{j=1}^N  s_j \left[ \sigma(r_j - r_i) + \sigma(r_j^2) \right]
\end{equation}
where $r_i$ is the score for the positive sample, $r_j$ are scores for negative samples, $s_j = \text{softmax}(r_j)$, and $\sigma$ is the sigmoid function.
The final loss function for the graph based recommendation module combines the above loss functions as $\mathcal{L}_{rec} = \mathcal{L}_{BCE} + \mathcal{L}_{TOP1-max}$

The above framework is used to rank the dimensions and expressions to monitor. The specifics for each are outlined below.

\subsubsection{Dimension Recommendation}
The monitor entity graph would only consist of monitor, metrics and dimension nodes. The node features would consist of the ontology describing the monitors, metrics, dimensions or learned node embeddings in the absence of text features. The edge types are 
{\tt (monitor, has, metric), (metric, has, dimension), (monitor, associated\_with, dimension)}
along with the reverse edges. The output is a ranked set of dimensions for the given (monitor,metric) of a certain account.

\subsubsection{Expression Recommendation}
The graph comprises of the nodes: service/account, monitor, metric, dimension, expression. The node features are the textual ontology and the edge types are 
{\tt (service, has, monitor), (monitor, has, metric), (monitor, associated\_with, dimension), (metric, has, dimension), (monitor/metric/dimension, uses, expression)} 
along with the reverse edges. The task is to obtain a ranked set of expressions for the given (monitor,metric,dimension).

\begin{figure}[t!]
\fbox{\begin{minipage}{0.98\columnwidth}\small
\vspace{2pt}
You are an expert service engineer. Your task is to design the configuration of the alert expressions and thresholds for the given service...\\
You will be given the service details such as the name of the service, the metric (time-series) name, the time series values (if available)... You will also be given similar alert expressions, thresholds and best practices from similar metrics.\\
Glossary of the operators and definitions with examples: ...\\
Given service information: \{Account:..., Metrics:..., Dimensions:..., Sampling Types:..., Raw Timeseries:..., Percentile data:..., best practices:...\} \\
Below are the alert conditions of similar metrics: ...\\
Please generate the alert conditions in the following format: {{...}}
\end{minipage}}
\centering
\caption{The structure of the LLM prompt used to obtain the final alert configurations from the recommended values.}
\label{fig:reasoning_module_prompt}
\vspace{-0.15in}
\end{figure}

\subsection{Generation of Alert Conditions}\label{alert_cond_gpt}
In the previous sections we have seen how we obtain the recommendations from each module namely: metrics, dimensions, expressions and similar metrics' alert conditions. Now we see how we combine the individual recommendations into one coherent alert condition for final consumption for the purpose of monitor creation. Note, since the monitor is not yet created we may not have timeseries data to set the thresholds. Moreover, even if there is some timeseries available from the corresponding metric, it may not capture all the anomalies in the early phases and so may not be a good indicator of the threshold as seen in \emph{observation 4}. Hence, rather than statistical methods, have to rely on similar monitors or domain knowledge to set the baseline threshold which is adapted later based on the metric streams. Thus we use an LLM to generate the alert condition in the desired format from the recommended components and historical alerts.
The rationale behind using LLMs for generating the alert thresholds is they are good at processing textual information to mine semantically similar monitors and understand documented best practice of respective services.
The role of the LLM here is two fold: i) helping decide the similar metrics from the textual descriptions and retrieve thresholds from best practices (as similar metrics have similar alert conditions, cf. \emph{observation 5}) ii) consolidating the alert conditions in the format as desired by the respective monitor (each monitor has a separate format which is prompted in context). The prompt used as well as the alert format is shown in the fig. \ref{fig:reasoning_module_prompt}.

\noindent \textbf{Computational complexity}: The time complexity for the metric selection module is $\mathcal{O}(\lvert V \rvert)$ and that for the graph based recommendation module is $\mathcal{O}(\lvert E \rvert)$, which is linear in number of nodes and edges respectively. 
The LLM modules in alert generation receive a constant number of similar metrics and monitors in the prompt and have a constant time complexity per recommendation.

%% file: Results.tex
\section{Results} \label{results_sec}
In this section, we evaluate the effectiveness of the unified framework using real-world datasets collected from \company, a large-scale cloud system provider.  In particular, we aim to answer the following RQs:
\begin{itemize}[leftmargin=*]
    \item RQ1: What is the effectiveness of the individual modules?
    \item RQ2: How effective is the proposed unified framework compared to baselines, for monitor configuration recommendation?
    \item RQ3: What is the importance of each module in the framework?
    \item RQ4: Are the generated alerts from the system, meaningful for services in production? 
\end{itemize}

\textbf{Implementation Details:}
We decide the parameters using the standard practice adopted by similar methods \cite{velickovic2017graph, collaborative_filtering_Wang_2021} and do not perform extensive hyperparameter tuning. For fair comparison all the baselines use the same setting and number of parameters where applicable. The metric selection module has 2 layers and in the recommendation modules, 3 MP layers are used with hidden channel  size, 1024, 256 and output channel size 256, 128 respectively. For contrastive learning we sample upto 10 negative samples for computational efficiency.
We use the Adam optimizer with learning rate of $10^{-3}$ and weight decay of $10^{-5}$. To stabilize the training we employ a learning rate scheduler that decreases the learning rate by a factor of 2 if validation metrics don't improve for 5 epochs. In order to prevent overfitting we further use early stopping if the validation results stagnate for 10 contiguous epochs. The data (links for recommendation modules) is split into 80\% for training, 10\% for validation and 10\% for testing. In the graph training data, 70\% of the edges are used for message passing and 30\% for supervision. Unless stated otherwise, the GPT-4o \cite{openai2024gpt4ocard} model is being used for the LLM.
Below we provide the results from individual components and combined framework along with human evaluations.

\subsection{Results of individual modules (RQ1)}

\textbf{Metric Selection}:
Table \ref{tab:metric_sel_results} shows the results on the metric selection problem. We report similar metrics as \cite{metric_paper_icse} (the only baseline for metric selection available) for fair comparison namely: accuracy, precision, recall, f1-score (micro), macro f1-score and the hamming loss. We can see that our learning based approach outperforms the SVD approach of \cite{metric_paper_icse} (p<0.05, Wilcoxon signed rank test). Moreover the method is computationally expensive requiring cubic complexity to compute the SVD. However our method requires no such preprocessing and only involves a forward pass through the network over the metrics requiring linear complexity. 

\begin{table}[h!]
  \caption{Metric selection results wrt the baseline \cite{metric_paper_icse}. We outperform the baseline while having lower computational complexity ($\mathcal{O}(N)$ vs $\mathcal{O}(N^3)$, $N$=No. of metrics).} 
    \label{tab:metric_sel_results}
    \centering
    \resizebox{\columnwidth}{!}{
    \begin{tabular}{c c c c c c c}
    \hline
      \textbf{Metric} & \textbf{Acc.}($\uparrow$) & \textbf{Prec.}($\uparrow$) & \textbf{Rec.}($\uparrow$) & \textbf{F1-score}($\uparrow$) & \textbf{Macro-F1}($\uparrow$) & \textbf{Hamming}($\downarrow$)\\
        \hline \hline
        SVD & 0.841 & 0.754 & 0.840 & 0.795 & 0.431 & 0.159\\
        Ours & 0.866 & 0.773 & 0.863 & 0.816 & 0.453 & 0.134\\
    \hline
    \end{tabular}
    }
\end{table}

\textbf{Dimension/Expression Recommendation}:
For the dimension (expression) recommendation, we evaluate the performance of our model using the standard ranking based evaluation metrics of Hit Ratio (HR@k), Mean Reciprocal Rank (MRR), Normalized Discounted Cumulative Gain (NDCG@k), Recall@k. We compare against traditional collaborative filtering \cite{collaborative_filtering_Wang_2021}, non-graph based MLP \cite{MLP_Rumelhart1986} (using only the text features), standard graph methods such as SAGEConv with mean/max pooling (SAGE(v1), SAGE(v2)) \cite{hamilton2017inductive}, GATConv (GAT) \cite{velickovic2017graph}, Transformer (T) \cite{shi2020masked} and the methods for heterogeneous graphs such as HGT \cite{HGT}, HAN \cite{HAN}, HetGNN \cite{HetGNN}. 
From the results presented in table \ref{tab:rec_results}, we observe that the MLP model outperforms the user based collaborative filtering indicating the usefulness of the text features. Further we note the best GNN baseline (SAGE-conv) performs better than the vanilla MLP over the features, demonstrating that incorporating the graph structure into the learning helps in addition to the node features.
We see that our model with ranking loss (+RL) drastically enhances the results over baselines (p<0.05, Wilcoxon test, deviations within 0.002).
Our attention based model coupled with the losses and the induced metapaths (+Paths) shows the best overall performance.
Similar results are observed for expression recommendation and for brevity we only show the dimension recommendation baseline results.

Thus overall the metric, dimension and expression recommendation modules are able to outperform the baseline and provide competitive results answering RQ1.

\begin{table}[h!]
  \caption{Proposed framework outperforms baseline graph and non-graph methods. The baseline results show only dimension rec. Similar results are obtained for expression rec.} 
    \label{tab:rec_results}
    \centering
    \resizebox{\columnwidth}{!}{
    \begin{tabular}{c c c c c c c c c}
    \hline
      \textbf{Metric} & \textbf{HR@1} & \textbf{HR@3} & \textbf{HR@5} & \textbf{MRR} & \textbf{N@k} & \textbf{R@1} & \textbf{R@3} & \textbf{R@5} \\
        \hline \hline
        CF & 0.230 & 0.149 & 0.116 & 0.391 & 0.217 & 0.117 & 0.310 & 0.438\\
        MLP & 0.312 & 0.186 & 0.128 & 0.464 & 0.307 & 0.184 & 0.392 & 0.492\\
        \hline
        SAGE(v1) & 0.383 & 0.186 & 0.127 & 0.499 & 0.328 & 0.218 & 0.379 & 0.474\\
        SAGE(v2) & 0.291 & 0.154 & 0.111 & 0.414 & 0.262 & 0.165 & 0.323 & 0.398\\
        GAT        & 0.355 & 0.185 & 0.18 & 0.487 & 0.323 & 0.213 & 0.397 & 0.493 \\
        HGT        & 0.396 & 0.185 & 0.131 & 0.510 & 0.356 & 0.228 & 0.402 & 0.507 \\
        HAN        & 0.348 & 0.194 & 0.131 & 0.485 & 0.315 & 0.202 & 0.407 & 0.500 \\
        HetGNN        & 0.375 & 0.173 & 0.121 & 0.492 & 0.321 & 0.224 & 0.400 & 0.502 \\
        T (dim. rec.) & 0.331 & 0.188 & 0.134 & 0.481 & 0.306 & 0.178 & 0.399 & 0.523 \\
        T (exp. rec.) & 0.241 & 0.386 & 0.466 & 0.349 & 0.246 & 0.188 & 0.292 & 0.362 \\
        \hline
         \makecell{T + RL (dim. rec.)} & 0.573 & 0.246 & 0.159 & 0.672 & 0.525 & 0.342 & 0.592 & 0.675 \\
         \makecell{+ Paths (dim. rec.)} & \textbf{0.597} & \textbf{0.265} & \textbf{0.173} & \textbf{0.714} & \textbf{0.555} & \textbf{0.355} & \textbf{0.649} & \textbf{0.748} \\
         \makecell{+ Paths (exp. rec.)} & \textbf{0.535} & \textbf{0.701} & \textbf{0.750} & \textbf{0.631} & \textbf{0.519} & \textbf{0.380} & \textbf{0.596} & \textbf{0.662} \\
         \hline
    \end{tabular}
    }
\end{table}

\subsection{Results of Unified Framework (RQ2, RQ3)}

\begin{figure}[h!]  
\centering
\begin{subfigure}[b]{0.22\textwidth}
       \centering
\includegraphics[width=\linewidth]{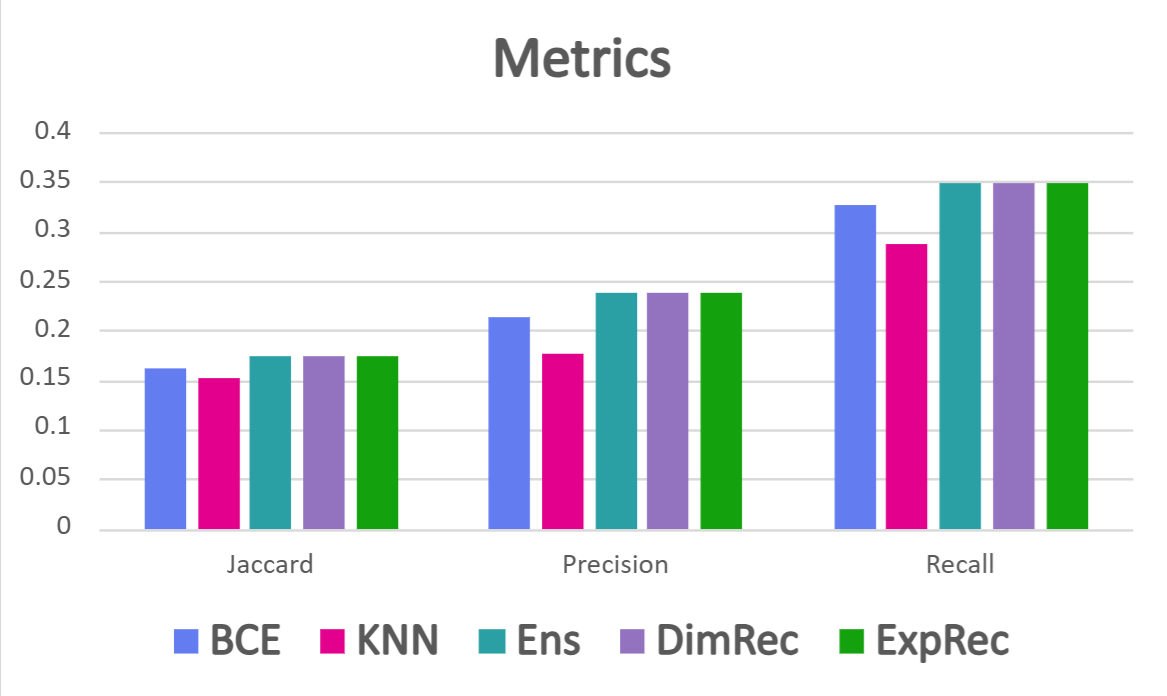}%
\caption{}
\label{subfig: main_results1}
\end{subfigure}
\begin{subfigure}[b]{0.22\textwidth}
       \centering
\includegraphics[width=\linewidth]{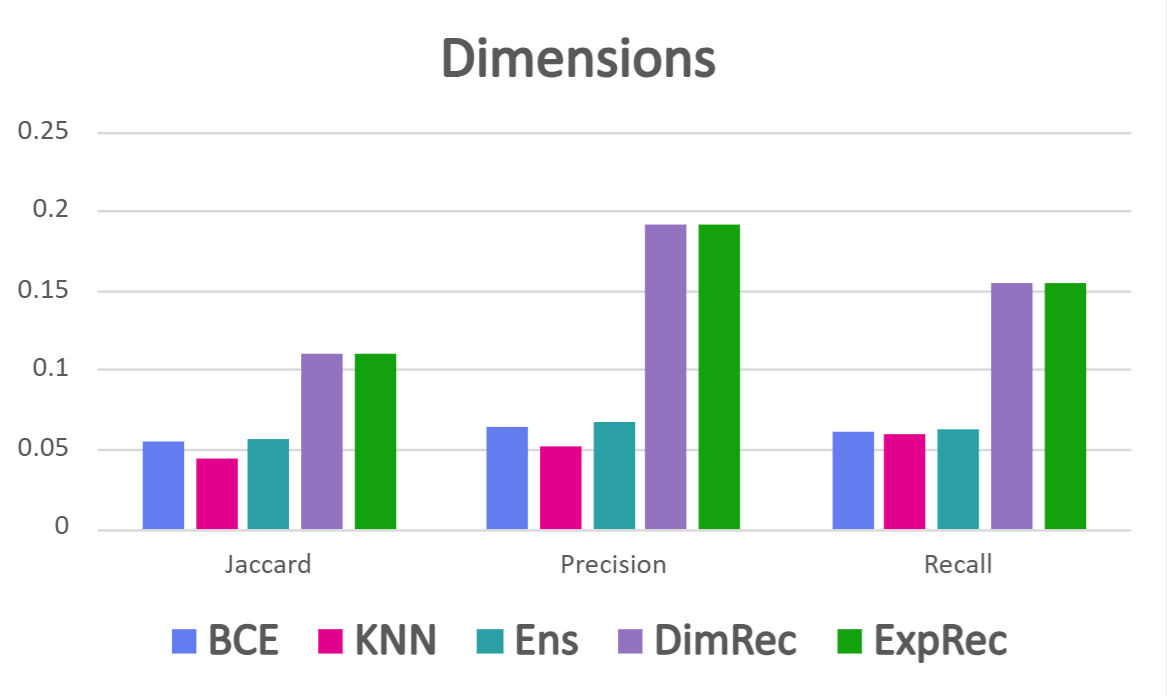}
\caption{}
\label{subfig: main_results2}
\end{subfigure}

\begin{subfigure}[b]{0.22\textwidth}
       \centering
\includegraphics[width=\linewidth]{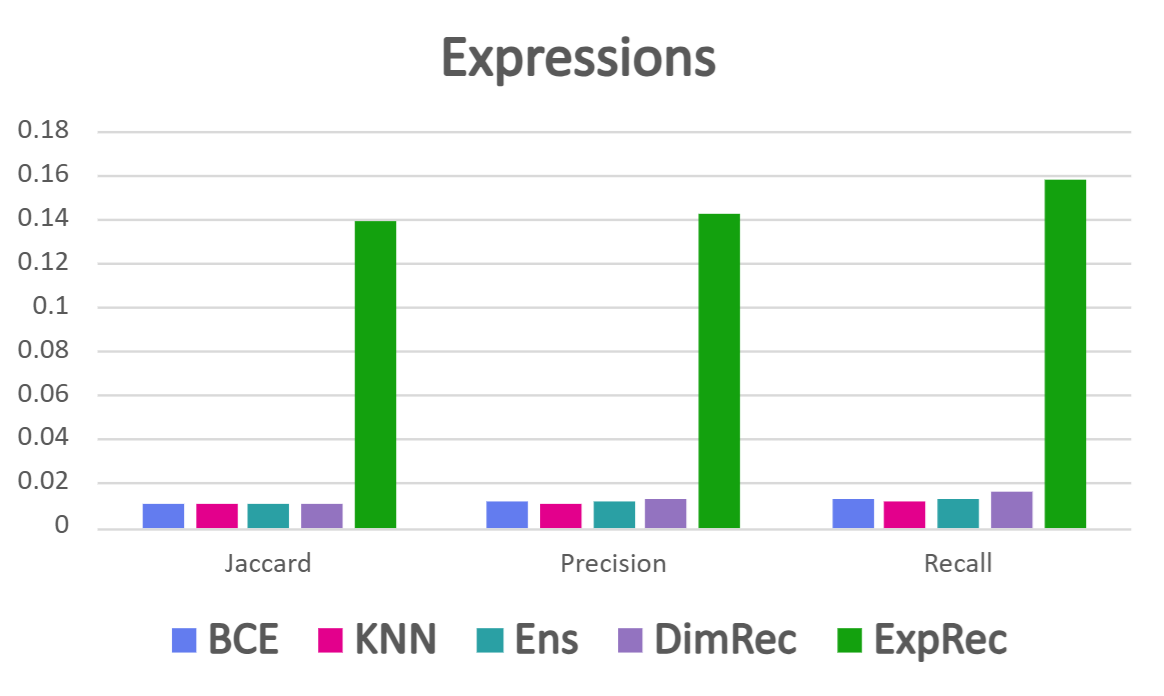}%
\caption{}
\label{subfig: main_results3}
\end{subfigure}
\begin{subfigure}[b]{0.22\textwidth}
       \centering
\includegraphics[width=\linewidth]{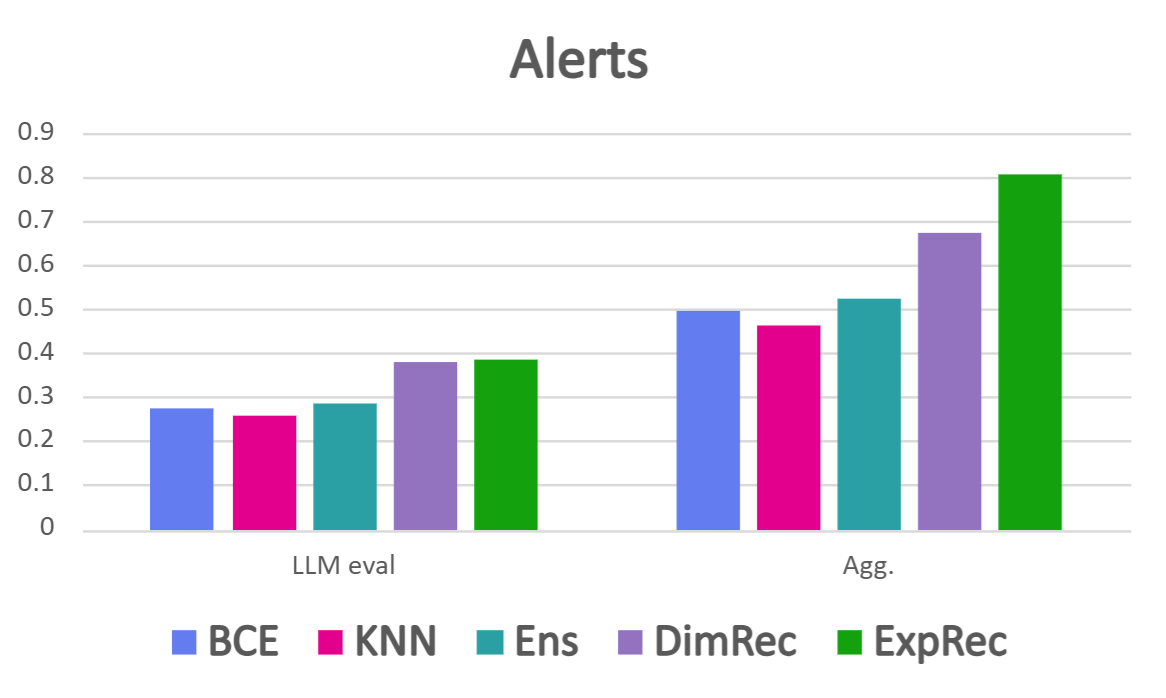}
\caption{}
\label{subfig: main_results4}
\end{subfigure}
    \caption{Results of the Unified Framework. The colors represent the model configuration used for system evaluation. For metric selection (a), we observe best performance by the ensemble (Ens) model. On further incorporating the graph based recommendation models (DimRec:Ens+DimRec, ExpRec:Ens+Dimrec+ExpRec) we find a drastic improvement in the individual and overall results (b,c,d).}
    \label{fig: main_results}
\end{figure}

In this section we present the evaluation results of the overall framework combining all the modules in the pipeline. Specifically the pipeline consists of the metric, dimension and expression recommendation modules and the module to combine these into the alert conditions. We evaluate the three variants of the metric recommendation as described in section \ref{metric_rec} namely: 1) BCE: using BCE loss 2) KNN: using the contrastive loss and 3) Ens: the ensemble combining the two objectives. In order to show the importance of the dimension and expression recommendation modules, we first run the pipeline without these modules obtaining the recommendations from the LLM. Then we incrementally add these modules to the best metric selection method (DimRec: Ens+DimRec, ExpRec: Ens+DimRec+ExpRec) and observe the results. For evaluation, we use the Jaccard similarity, precision and recall for the recommendation modules. For alerts, we use an LLM evaluation framework where we provide all the text/timeseries features of the account, the expected \& predicted alerts and ask the LLM 
to provide a rating between 0 and 1 of whether the predicted alerts are able to detect the production incidents.
Since alerts are functions, comparing them is not as straightforward as comparing strings (in previous modules). Therefore, we approach the evaluation by systematically prompting LLMs using the prompt structure in \autoref{fig:llm_eval_prompt} and leave comparison by actual incident evaluation using the alerts for future work. Nevertheless, to ground the results, we correlate the LLM evaluation with a human evaluation in the next section. The results are presented in \autoref{fig: main_results},
where each bar in a cluster represents results from a system run using corresponding model configuration. 
We see that the BCE variant of the metric recommendation works better than the contrastive one (perhaps as we do not perform much optimizations). 
The joint optimization method (Ens) combining the two performs best. We use this for further experiments and notice that as we add the dimension recommendation module the results (for dimension and overall) improve. Further when the expression recommendation module is added the performance is enhanced indicating the importance of each module answering RQ2, RQ3. All improvements are statistically significant (p<0.05, Wilcoxon test).

\begin{figure}[h!]
\fbox{\begin{minipage}{0.98\columnwidth}\small
\vspace{2pt}
You are evaluating alert recommendations for a cloud monitoring system.

Account Context: \{Service: ..., Metrics: ..., ...\}

Alert Conditions to evaluate: ...

Actual Alert Conditions Set: ...

1st percentile of the metrics' timeseries data: ...

99th percentile of the metrics' timeseries data: ...

Evaluation Criteria:
Rate each criteria from [0,1] (higher the better):

1. \underline{Threshold} Appropriateness: ...
2. \underline{Condition} Validity: ...
3. Detecting \underline{Customer Incidents}: ...
4. \underline{Noise} reduction: ...
5. \underline{Specificity}: ...
6. \underline{Completeness} of alerts: ...
7. \underline{Checks} for required fields: ...

Response Format: \{JSON response format\}
\end{minipage}}
\centering
\caption{The structure of the LLM prompt used to evaluate the expected and generated alert conditions. Final LLM eval score is the aggregate of the scores for the individual criteria.}
\label{fig:llm_eval_prompt}
\vspace{-0.15in}
\end{figure}

\subsection{Human Evaluation and User Study (RQ4)}
In order to ground the results of the LLM evaluation in the previous section, we evaluate the alerting conditions by human raters (3 SMEs). Specifically, we have the SMEs review the predicted and expected alerts for $\sim 100$ samples and give a score from [0,1]. A perfect score of 1 indicates all the expected alerts were captured by the recommendations and a score of 0 means none of the predicted alerts reflected the expected alerting conditions. An inter rater agreement is computed between the LLM evaluation score and the human provided scores (majority vote) using the Cohen's kappa \cite{cohens_kappa}. We find a score of 0.62 indicating a significant rating agreement between the LLM and the human. 
Further we find a statistically significant (p value $< 10^{-4}$) correlation of 0.673 between the two scores (avg. of human score. 
The high inter rater agreement and correlations show that the LLM evaluation framework is indeed indicative of the expected results at scale  
answering RQ4.

\textbf{User Study and lessons learnt for production services:}
We perform a study on a subset of users (sample set $\approx 30$) from product teams (service owners) who use our recommendations. The aim of the study was to find out the major challenges encountered while creating and updating monitors, whether the recommendations helped and  where the provided recommendations could be improved. It was observed that most of the services update monitors at a frequency of one week. The major challenges faced were determining the subset of dimensions to monitor on and the thresholds at which to raise an  alert. 
We also learnt that \emph{recommendations were more actionable if we provide explanations using similar services, reflecting the way configurations are set manually}. This informs our alert generation module that uses best practices from similar services.
In addition, one user explicitly called out the need for an end-to-end automation of monitor creation and also surfacing the recommendations using prompt boxes. Overall the users perceived the recommendations as helpful during creation and while updating the monitors, with an average relevance rating of 4.5 out of 5.

\section{Threats to Validity}

\textbf{Generalizability}
Our empirical study and framework evaluation are based on data from \company and the results may not hold elsewhere. The framework’s reliance on historical monitor configurations and company-specific ontologies may limit its applicability elsewhere. However, our experiments cover a diverse and extensive dataset spanning 10's of thousands of services and monitors. Future work should evaluate the approach in other industrial contexts.

\noindent \textbf{Construct Validity}
The evaluation of the proposed framework relies on a combination of standard metrics (accuracy, precision, recall, F1-score, NDCG, etc.) and LLM-based scoring for alert condition quality. While these metrics are widely used, they may not fully capture the practical effectiveness of monitor recommendations in real-world production environments. The LLM-based evaluation, in particular, may not reflect all nuances of incident detection and alerting in practice. To address this, we complemented LLM-based assessments with human SME evaluations and reported the inter-rater agreement which aligns with the LLM evaluation. We further gather feedback from production service owners who affirm the relevancy of the recommendations.

\section{Discussion}

To the best of our knowledge, this paper is the first to propose and layout a framework for recommending configurations for monitoring services in a large scale cloud setting. 
Following are some future research directions to pursue:

\noindent \textbf{1) Explicit usage of incident data: } The current system implicitly makes use of incident data in the form of monitors created from historical incidents. However, we do not yet use the incident logs, discussions and proposed resolutions into the monitor creation process explicitly. Incorporating these signal could help in creating monitors as and when the incidents occur.

\noindent \textbf{2) Principled system evaluation: } 
In this paper we have evaluated the final alert conditions using LLMs and verified with human evaluation. A more principled approach would be to evaluate these alerting conditions in the actual cloud environment.

\noindent \textbf{3) Continuous improvement: } 
A challenging future research topic would be to design an autonomous system that gathers implicit feedback from customer reported incidents, incident logs, discussions etc. and improves in a weakly supervised manner. 

\noindent \textbf{4) Towards self-healing cloud: } This paper focuses on obtaining configurations for accurate incident detection. However, once the incident is detected a human has to spend time in root causing and resolving the issue. It would be an interesting research direction to pursue whether the intelligent process of detecting incidents could help with diagnosing and further mitigating the incident.

%% file: Conclusion.tex
\section{Conclusion}

In this paper, we propose a holistic and end-to-end framework for recommending monitor configurations in a large scale cloud environment. We study the monitoring process for production services in \company and design recommendation systems to obtain predictions for individual components. We further build a system combining the individual components, to provide coherent configuration recommendations.
The proposed evaluation framework shows that the method performs competitively in recommending monitor configurations corroborated by human evaluation. We hope this work inspires future research in intelligent monitoring and incident detection of cloud services.